\documentclass{patera}

\usepackage[cp1251]{inputenc}
\def\C{\mathbb C}
\def\R{\mathbb R}
\def\Z{\mathbb Z}
\def\N{\mathbb N}

\def\r{\rangle}
\def\l{\langle}

\def\w{\omega}

\def\r{\rangle}
\def\l{\langle}

\newcommand{\comment}[1]{}

\newtheorem{theorem}{Theorem}

\newtheorem*{lemma*}{Lemma}
\newtheorem{corollary}{Corollary}
\newtheorem{proposition}{Proposition}
\newtheorem*{conjecture*}{Conjecture}

{\theoremstyle{definition}
\newtheorem{definition}{Definition}
\newtheorem{example}{Example}
\newtheorem{remark}{Remark}

\newtheorem*{note*}{Note}
}

\allowdisplaybreaks

\numberwithin{equation}{section}

\begin{document}

\begin{flushleft}
\LARGE \bf Orthogonal polynomials of compact simple Lie groups
\end{flushleft}

\begin{flushleft} \it
Maryna NESTERENKO~$^\dag$,
Ji\v{r}\'{i} PATERA~$^\ddag$
and
Agnieszka TERESZKIEWICZ~$^\S$
\end{flushleft}

\noindent $^\dag$~Institute of Mathematics of NAS of Ukraine, 3 Tereshchenkivs'ka Str., Kyiv-4, 01601 Ukraine

\noindent $^\ddag$~Centre de recherches math\'ematiques, Universit\'e de Montr\'eal, C.P.6128-Centre ville, Montr\'eal,\\
\phantom{$^\ddag$}~H3C\,3J7,~Qu\'ebec,~Canada

\noindent $^\S$~Institute of Mathematics, University of Bialystok, Akademicka 2, PL-15-267 Bialystok, Poland

\noindent $\phantom{^\S}$~E-mail: maryna@imath.kiev.ua, patera@crm.umontreal.ca, a.tereszkiewicz@uwb.edu.pl

{\vspace{9mm}\par\noindent\hspace*{8mm}\parbox{140mm}
{\small
Recursive algebraic construction of two infinite families of polynomials in $n$ variables is proposed as a uniform method applicable to every semisimple Lie group of rank~$n$.
Its result recognizes Chebyshev polynomials of the first and second kind as the special case of the simple group of type $A_1$.
%It is done by a substitution of variables into the known functions formed by summing up exponential terms over orbits of the Weyl groups of the Lie group.
%Real variables are replaced by the lowest orbit functions as the new variables.
The obtained not Laurent-type polynomials are equivalent to the partial cases of the Macdonald symmetric polynomials.
%Basic relations between the polynomials and their properties follow from the corresponding properties of the orbit functions, namely the orthogonality and discretization.
Recurrence relations are shown for the Lie groups of types $A_1$, $A_2$, $A_3$, $C_2$, $C_3$, $G_2$, and $B_3$ together with lowest polynomials.

\vspace{3mm}

}
\par\vspace{3mm}}

%%%%%%%%%%%%%%%%%%%%%%%%%%%%%%%%%%%%%%%%%%%%%%%%%%%%%%%%%%%%%%
\section{Introduction}\label{Introduction}
%%%%%%%%%%%%%%%%%%%%%%%%%%%%%%%%%%%%%%%%%%%%%%%%%%%%%%%%%%%%%
The majority of special functions and orthogonal polynomials  introduced during the last decade
are associated with Lie groups or their generalizations.
In particular, special functions of mathe\-ma\-ti\-cal physics are in fact matrix elements of representations of Lie groups~\cite{VilenkinKlimyk1995}
and recent multivariate generalizations of classical hypergeometric orthogonal polynomials are based on root systems of simple Lie
groups/algebras~\cite{Dunkl,HeckSchlich, Koornwinder1993, Mac1, Mac2, Mac3,Opdam,vanDiejen2}.
In this connection a~number of elegant results in theory these families of orthogonal polynomials,
such as explicit (determinantal) computation of polynomials~\cite{Lapointe1,Lapointe2,vanDiejen1} and
Pieri formulas~\cite{Lassalle2006, vanDiejen3}, were obtained, see also~\cite{Lassalle, LassalleSchlosser2010,Kowalski1982} and references therein.

The main purpose of this article is to construct orthogonal polynomials in $n$ variables based on orbit functions related to simple Lie groups of rank $n$ (see Section~\ref{ssec_orb_func_char} for definitions).
As far as we deal with the functions invariant/skew-invariant under the action of the corresponding Weyl group
the obtained polynomials appear as building blocks in all multivariate polynomials associated with root systems.
Unlike Gram--Schmidt type orthogonalization of the monomial basis with respect to Haar measure~\cite{HeckSchlich, Mac3, Opdam}
or determinantal construction of polynomials~\cite{Lapointe1,Lapointe2,vanDiejen1}
we make profit from decomposition of products of Weyl group orbits and from basic properties of the characters of irreducible finite dimensional representations.

Relating the polynomials to the Lie groups, by means of either the characters or closely related to it the Weyl group orbit functions ($C$- and $S$-functions),
allows one to carry over power\-ful results of the accomplished theory of the compact simple Lie groups as properties of the polynomials.
Let us point out the following properties:

\begin{itemize}\itemsep=0pt
\item[(a)]
The domain of orthogonality $F$ of characters and/or orbit functions coincide in the real Euclidean space $\R^n$.
The domain $F$ is known \cite{Bourbaki} for all $G$.
The polynomial substitution of variables transforms $F$ into the domain of orthogonality $\tilde F$ for the polynomials.
Examples of rather bizarre shape of $\tilde F$ of the  Lie groups $C_2$ and  $G_2$  are shown in~\cite{HMP} and~\cite{MMP} respectively.

\item[(b)]
Orthogonality of characters, $C$-, and $S$-functions, when integrated over $F$, is known~\cite{MP06}.
From it follows directly the orthogonality of the polynomials when integrated over $\tilde F$.

\item[(c)]
Discretization and discrete  orthogonality of characters and $C$-functions of \cite{MP87} extends directly to $S$-functions~\cite{MP06}.
For any $G$ it is done on the fragment of a lattice in $F$ of any chosen density.
Lattice points in $F$ become discrete set of points in $\tilde F$  after the polynomial substitution of variables.

\item[(d)] Congruence classes of characters, orbit functions and of the polynomials are a practically useful in computing with these objects~\cite{LP}.
They arise from the action of the center of the corresponding compact simple Lie group.
\end{itemize}

Other useful features of orbit functions~\cite{KlimykPatera2006, KlimykPatera2007-1} should be reflected in properties of polynomials originating from simple Lie groups: in particular, orbit functions are eigenfunctions of differential operators, the Laplace operator being one of them.
$C$- and $S$-functions are solutions of the Neumann and Dirichlet boundary value problem respectively.
However, our primary objective at this stage is to establish a constructive method for finding orthogonal multivariate polynomials,
indeed, for actually seeing them.

Our method is purely algebraic and we propose three different ways to transform a $C$- or $S$-orbit function into a polynomial.

The first one substitutes for each multivariable exponential term in an orbit function a monomial of as many variables~\eqref{subst_exp}.
In~$1D$ this results in Chebyshev polynomials written as Laurent polynomials with symmetrically placed positive and negative powers of the variable;
and in the case of $A_2$ our results coincide with those from~\cite{Koornwinder1-4}.

The~second method, the `truly trigonometric' form, is based on the fact that, for many simple Lie algebras (see the list in~\eqref{list} below),
each $C$ and $S$-orbit function consists of pairs of exponential terms that add up to either cosine or sine.
Hence such a function is a sum of trigonometric terms.
For the Chebyshev polynomials we obtain in this way their trigonometric form.
Note from~\eqref{list} that this method does no apply to the groups $A_n$ for $n>1$.

This paper focuses on polynomials obtained by the third substitution of variables,
mimicking Weyl's method for the construction of finite dimensional representations from $n$ fundamental representations (see for example~\cite{dynkin}, Supplement, Section~5).
Thus the $C$-polynomials have $n$ variables that are the $C$-orbit functions, one for each fundamental weight $\w_j$, see~\eqref{bases}.
%$S$-polynomials differ from the characters by an additional unique variable, the lowest $S$-function denoted~$S_\rho$.
This approach results in a simple recursive construction that allows one to represent any orbit function/monomial symmetric function in non Laurent polynomial form.

In addition to the general approach and associated tools we present a lot of explicit and practically useful data and discussions,
namely in Appendix~\ref{sec_A_1} we compare the classical Chebyshev polynomials (Dickson polynomials)
and orbit functions of $A_1$ with their recursion relations.
Not quite standard is addition of the $A_1$-character formula~\eqref{A1character}, linking $C$- and $S$-polynomials.
Suitably normalized, the Chebyshev polynomials of the first and second kind coincide with the $C$- and $S$-polynomials.
A table of the polynomials of each kind is presented.
Appendices~\ref{sec_C_2}, \ref{sec_A_2}, and \ref{sec_G_2}
contain respectively the recursion relations for polynomials of the Lie algebras $A_2$, $C_2$ and  $G_2$.
In Appendix~\ref{sec_3D} recursion relations for $A_3$ and generic recursion relations for $B_3$ and $C_3$
polynomials of both kinds are listed together with useful tools for solving these recursion relations,
namely the formulas determining the congruence class of polynomials and the dimensions of the irreducible representations.

%%%%%%%%%%%%%%%%%%%%%%%%%%%%%%%%%%%%%%%%%%%%%%%%%%%%%%%%%%%%%
\section{Preliminaries and conventions}
%%%%%%%%%%%%%%%%%%%%%%%%%%%%%%%%%%%%%%%%%%%%%%%%%%%%%%%%%%%%%
This section serves to fix notations and terminology and to recall the definitions and some of the properties of orbit functions. Additional details can be found for example in~\cite{Bourbaki,HrivnakPatera2009,Humphreys1972,KassMoodyPateraSlansky1990,
KlimykPatera2006, KlimykPatera2007-1, KlimykPatera2008, KlimykPatera2007-3}.

%%%%%%%%%%%%%%%%%%%%%%%%%%%%%%%%%%%%%%%%%%%%%%%%%%
\subsection{Notations}
Let $\R^n$ be the Euclidean space spanned by the simple roots
of a~simple Lie group~$G$. The basis of the simple roots and the basis of fundamental weights are hereafter referred to as the $\alpha$-basis and $\omega$-basis respectively. The two bases are linked by the Cartan matrix $M$ of $G$. In matrix form that~is
\begin{gather*}%\label{bases}
\alpha=M\omega,\qquad
\omega=M^{-1}\alpha,\qquad
M=(M_{jk})=
\left(\frac{2\langle\alpha_j,\alpha_k\rangle}{\langle\alpha_k,\alpha_k\rangle}\right),\qquad
j,k=\{1,2,\dots,n\}.
\end{gather*}

Bases dual to $\alpha$- and $\omega$-bases are denoted
by $\check{\alpha}$- and $\check{\omega}$-bases. In addition one uses
$\{e_1,\ldots, e_n\}$, the orthonormal basis of $\R^n$.
Note, that
$\langle\alpha_j,\check{\omega}_k\rangle=\langle\check{\alpha}_j,\omega_k\rangle=\langle e_j,e_k\rangle=\delta_{jk}$,
where $\langle\cdot\,,\cdot\rangle$ is the inner product and $\delta_{jk}$ is the Kronecker delta.

The root lattice $Q$ and the weight lattice $P$  of $G$ are formed
by all integer linear combinations of the $\alpha$-basis and
$\omega$-basis,
\begin{gather}\label{bases}
Q=\Z\alpha_1+\Z\alpha_2+\cdots+\Z\alpha_n,\qquad
P=\Z\omega_1+\Z\omega_2+\cdots+\Z\omega_n.
\end{gather}
In the weight lattice $P$, we define the cone of dominant weights
$P^+$ and its subset of strictly dominant weights $P^{++}$
\begin{gather*}
P\;\supset\; P^+=\Z^{\ge 0}\omega_1+\cdots+\Z^{\ge 0}\omega_n
\;\supset\; P^{++}=\Z^{>0}\omega_1+\cdots+\Z^{>0}\omega_n.
\end{gather*}

Hereafter $W=W(G)$ is the Weyl group, i.e., the finite group generated by reflections in $(n-1)$-dimensional hyperplanes orthogonal to simple roots, having the origin as their common point, and referred to as elementary reflections $r_j$, $j=1,\dots, n$.
The orbit of $W$ containing the (dominant) point $\lambda\in P^+\subset \R^n$ is written as  $W_\lambda$.
The size of $W_\lambda$ is denoted by  $|W_\lambda|$, it is the number of points in $W_\lambda$.

The fundamental region $F(G)\subset \R^n$ is the convex hull of the vertices
$\{0,\frac{\omega_1}{q_1},\ldots, \frac{\omega_n}{q_n}\}$,
where~$q_j$, $j=\overline{1,n}$ are comarks of the highest root $\xi$,
i.e., \mbox{$\xi=q_1\check{\alpha}_1+\dots+q_n\check{\alpha}_n$}.

%%%%%%%%%%%%%%%%%%%%%%%%%%%%%%%%%%%%%%%%%%%%%%%%%%%%%%
\subsection{Orbit functions and the character}\label{ssec_orb_func_char}
%%%%%%%%%%%%%%%%%%%%%%%%%%%%%%%%%%%%%%%%%%%%%%%%%%%%%%
An orbit function of $n$ variables is the set of distinct points in $\R^n$ generated by the action of $W(G)$ on $\lambda$.
\begin{definition}\label{C}
The $C$-function $C_{\lambda}(x)$ is defined as
\begin{gather*}\label{def_c-function1}
C_\lambda(x) := \sum_{\mu\in W_\lambda(G)} e^{2\pi i \l\mu, x\r},
\qquad x\in\R^n,\quad \lambda\in P^+.
\end{gather*}
\end{definition}

\begin{definition}
The $S$-function $S_{\lambda}(x)$ is defined as
\begin{gather*}\label{def_s-function1}
S_\lambda(x) := \sum_{\mu\in W_\lambda(G)} (-1)^{p(\mu)}e^{2\pi i\l\mu,x\r},
\qquad x\in\R^n,\quad \lambda\in P^{++}\,.
\end{gather*}
where $p(\mu)$ is the number of elementary reflections  necessary to obtain $\mu$ from~$\lambda$.
\end{definition}

The same $\mu$ can be obtained by different successions of reflections,
but all shortest routes from $\lambda$ to $\mu$ will have a length of the same parity, so $S$-functions are well defined.

%\begin{definition}
%An $S$- and $C$-polynomial, defined by one orbit of the Weyl group $W$, is referred to as irreducible $S$-, $C$-polynomial.
%\end{definition}

In this paper, we always suppose that $\lambda, \mu\in P$ are given in $\w$-basis and $x\in\R^n$ is given in $\check{\alpha}$-basis, namely
$\lambda=\sum\limits^n_{j=1}\lambda_j\w_j$,
$\mu=\sum\limits^n_{j=1}\mu_j\w_j$, $\lambda_j, \mu_j\in\Z$ and
$x=\sum\limits^n_{j=1}x_j\check{\alpha}_j$,   $x_j\in \R$.
Hence the orbit functions have the following forms
\begin{gather}
C_\lambda(x)
= \sum_{\mu\in W_\lambda} e^{2\pi i \sum\limits^n_{j=1}\mu_jx_j}
= \sum_{\mu\in W_\lambda} \prod\limits^n_{j=1} e^{2\pi i \mu_jx_j},
\label{Cdef}
\\
S_\lambda(x)
= \sum_{\mu\in W_\lambda} (-1)^{p(\mu)}e^{2\pi i \sum\limits^n_{j=1}\mu_jx_j}
= \sum_{\mu\in W_\lambda} (-1)^{p(\mu)}\prod\limits^n_{j=1} e^{2\pi i \mu_jx_j}\label{Sdef}.
\end{gather}

The number of exponential summands that form an irreducible $S$-function $S_\lambda$ is always equal to the order of the corresponding Weyl group, because the function is non-zero only if the stabilizer of $\lambda$ in $W$ is the identity element. The number of exponential summands that form an irreducible $C$-function $C_\lambda$ is equal to the order of $W$ divided by the order of the stabilizer of $\lambda$ in $W$. Note that in the 1-dimensional case, $C$- and $S$-functions are respectively cosine and sine functions up to a normalization. The special case, $C_{(0,0,\dots,0)}=1$ for any $G$, occurs if $\lambda$ is the origin of $\R^n$.

Occasionally it is useful to scale up $C_\lambda$ of non-generic $\lambda$ by the stabilizer of $\lambda$ in $W$.
It is done by somewhat modifying the Definition~\ref{C}.
Rather than summation over the elements of $W_\lambda$, one should sum over all the elements of $W$.
Then $\lambda$ would be counted as many times as there are elements  $w\in W$  stabilizing $\lambda$.
That is the elements with the property $w\lambda=\lambda$.
%The alternate definition then is
%\begin{definition}\label{C/alt}
%The $C$-function $C_{\lambda}(x)$, $\lambda\in P^+$ is defined as
%\begin{gather*}\label{def_c-function1}
%C_\lambda(x) := \sum_{w\in W(G)} e^{2\pi i \l w\lambda, x\r},
%\qquad
%x\in\R^n.
%\end{gather*}
%\end{definition}
%Computation of points of many specific $W$-orbits is found, for example, in \cite{HLP}.
%\medskip

There is a fundamental relation between the $C$ and $S$-orbit functions for simple Lie group~$G$ of any type and rank, called the Weyl character formula. For the character $\chi_\lambda(x)$ of the representation of $G$ with the highest weight $\lambda$, it is the following:
\begin{gather}\label{thecharacter}
\chi_\lambda(x)=\frac{S_{\lambda+\rho}(x)}{S_\rho(x)}
 =\sum_\mu m^\lambda_\mu C_\mu(x),\qquad
            x\in\R^n,\quad \lambda,\mu\in P^+.
\end{gather}
Here $\rho$ is the half sum of the positive roots of $G$, which is known to be for any $G$ given by
$\rho=\sum\limits_{k=1}^n\w_k$.
The positive integer $m^\lambda_\mu$ is the Kostka number (the multiplicity of the dominant weight~$\mu$ in the weight system of the irreducible representation with the highest weight $\lambda$).
A basic algorithm in representation theory \cite{MP82} is used for computation of multiplicities.
The multiplicity of the highest term in~\eqref{thecharacter} is always one, i.e.\ $m^\lambda_\lambda=1$ in all cases. A suitable ordering of the weights $\lambda\in P^+$ makes the matrix $(m^\lambda_\mu)$ triangular (see Tables in~\cite{BMP}) with all diagonal entries equal to one. Such a matrix can be easily inverted, so that any $C_\mu(x)$ can be written as a linear combination of irreducible characters with coefficients that are integer but not all positive.
\begin{gather}\label{thecharacterinv}
C_\mu(x)=\sum_\lambda (m^\lambda_\mu)^{-1} \chi_\lambda(x)=S^{-1}_\rho\sum_\lambda (m^\lambda_\mu)^{-1} S_{\lambda+\rho},\qquad
x\in\R^n,\quad \lambda,\mu\in P^+,
\end{gather}
where $(m^\lambda_\mu)^{-1}$ are elements of the triangular matrix inverse to $(m^\lambda_\mu)$.
%%%%%%%%%%%%%%%%%%%%%%%%%%%%%%%%%%%%
\begin{example}
This example illustrates the relations \eqref{thecharacter} and \eqref{thecharacterinv} between polynomials~$C$ and~$\chi$ of the Lie group/Lie algebra of type $G_2$.

Tables~\ref{G2ex} contain values of $m^\lambda_\mu$ and $(m^\lambda_\mu)^{-1}$ for the five lowest dominant weights of~$G_2$.
It is thus possible to write any of the lowest five characters as the linear combination of $C$-polynomials~(\ref{thecharacter}), as well as the inverse relation, namely writing $C$-polynomials in terms of the characters~(\ref{thecharacterinv}).
In~particular, using the entries in the last column of the first table, we obtain
\begin{gather}\label{chiex}
\chi_{(1,1)}=4C_{(0,0)}+4C_{(0,1)}+2C_{(1,0)}+2C_{(0,2)}+C_{(1,1)}.
\end{gather}
Similarly according to  (\ref{thecharacterinv}), one reads the last column of the second table as follows,
\begin{gather}\label{Cex}
C_{(1,1)}=2\chi_{(0,0)}-2\chi_{(0,2)}+\chi_{(1,1)}.
\end{gather}
A useful verification: The equalities in~\eqref{chiex} and~\eqref{Cex} must be maintained when $\chi_\lambda$ and $C_\mu$ are replaced by the dimensions of the representation (see~\eqref{dimG2}) and by the size of the W-orbits (see Table~\ref{G2ex}) respectively.
%%%%%%%%%%%%%%%%%%%%%%%%%%%%%%%%%%%%
\begin{table}\centering
\begin{tabular}{|@{\;}c@{\;}|@{\;}c@{\;}|@{\;}c@{\;}|@{\;}c@{\;}|@{\;}c@{\;}|@{\;}c@{\;}|
c@{\;}|c@{\;}|@{\;}c@{\;}|@{\;}c@{\;}|@{\;}c@{\;}|@{\;}c@{\;}|@{\;}c@{\;}|c@{\!}|c@{\;}|}
\cline{1-6}\cline{8-15}
&&&&&&\qquad\qquad&&&&&&&&\\[-8 pt]
$m^\lambda_\mu$ &(0,0)  &(0,1) &(1,0) &(0,2) &(1,1) &&$\big(m^\lambda_\mu\big)^{\!-\!1}$ &(0,0) &(0,1) &(1,0) &(0,2) &(1,1) &&$|W_\mu|$\\
&&&&&&\quad\quad&&&&&&&&\\[-10 pt]
\cline{1-6}\cline{8-13}\cline{15-15}
(0,0)            &1     &1     &2     &3     &4     &&(0,0)                  &1     &$-1$    &$-1$    &0    &2   &&1\\
\cline{1-6}\cline{8-13}\cline{15-15}
(0,1)            &      &1     &1     &2     &4     &&(0,1)                  &      &1     &$-1$    &$-1$   &0   &&6\\
\cline{1-6}\cline{8-13}\cline{15-15}
(1,0)            &      &      &1     &1     &2     &&(1,0)                  &      &      &1     &$-1$   &0   &&6\\
\cline{1-6}\cline{8-13}\cline{15-15}
(0,2)            &      &      &      &1     &2     &&(0,2)                  &      &      &      &1    &$-2$  &&6\\
\cline{1-6}\cline{8-13}\cline{15-15}
(1,1)            &      &      &      &      &1     &&(1,1)                  &      &      &      &     &1   &&12\\
\cline{1-6}\cline{8-13}\cline{15-15}
\multicolumn{6}{|c|}{\;}&&\multicolumn{8}{|c|}{\;}\\[-9 pt]
\cline{1-6}\cline{8-13}\cline{15-15}
dim              &1     &7     &14    &27    &64    &&dim                    &1     &7     &14    &27   &64  &&\\
\cline{1-6}\cline{8-15}
\end{tabular}
\caption{Tables of  multiplicities $m^\lambda_\mu$ and $(m^\lambda_\mu)^{-1}$ for the lowest 5 dominant weights of~$G_2$.
The~first row contains $\lambda$, the first column of each table contains $\mu$.
The~last column contains the size of the Weyl group orbit of the dominant weight $\mu$.
The last row contains the dimensions of the irreducible representations of $G_2$ with the highest weight $\lambda$.}\label{G2ex}
\centering
\end{table}
%%%%%%%%%%%%%%%%%%%%%%%%%%%%%%%%%%%%
\end{example}

%%%%%%%%%%%%%%%%%%%%%%%%%%%%%%%%%%%%%%%%%%%%%%%%%%%%%%
\subsection{Properties of orbit functions}
%%%%%%%%%%%%%%%%%%%%%%%%%%%%%%%%%%%%%%%%%%%%%%%%%%%%%%
The rank of the underlying semisimple Lie group/algebra is the number of variables of the three families of orbit functions.
In general, $C$- and $S$-functions are finite sums of exponential functions, therefore they are continuous and have continuous derivatives of all orders in $\R^n$. $S$-functions are antisymmetric with respect to the $(n-1)$-dimensional boundary of~$F$. Hence they are zero on the boundary of~$F$. $C$-functions are symmetric with respect to the $(n-1)$-dimensional boundary of~$F$. Their normal derivative at the boundary is equal to zero (because the normal derivative of a $C$-function is an $S$-function). Symmetry and other properties of the orbit functions are reviewed in \cite{KlimykPatera2006,KlimykPatera2007-1,KlimykPatera2008}.

Through the use of a method of orbit functions it is possible to solve eigenvalue problems
on the fundamental domain for any compact simple Lie group.
Namely it was shown~\cite{KlimykPatera2006, KlimykPatera2007-1} that $C_\lambda(x)$- and $S_\lambda(x)$-orbit functions are
eigenfunctions of the $n$-dimensional Laplace operator on the
simplexes, which are fundamental domains of compact simple Lie
groups, with the Neumann or Dirichlet boundary value conditions.
%The eigenvalues are given explicitly.
The Laplace operator has the same eigenvalues for every exponential function summand of an orbit function,
the eigenvalue equals to $-4\pi\langle \lambda,\lambda\rangle$.
Indeed, all points of the orbit containing $\lambda$ are equidistant from the origin.

For any complex square integrable functions $\phi(x)$ and $\psi(x)$, we define a continuous scalar product
\begin{gather}\label{def_cont_scalar_product}
\l\phi(x),\psi(x)\r:=\int_{{F}}\phi(x)\overline{\psi(x)}{\rm d}x.
\end{gather}
Here, integration is carried out with respect to the Euclidean measure, the bar means complex conjugation, and $x\in {F}$, where ${F}$ is the fundamental region of $W$.

Any pair of orbit functions from the same family is orthogonal with respect to the scalar product~(\ref{def_cont_scalar_product}) on the corresponding fundamental region \cite{HrivnakPatera2009}, namely
\begin{gather}\label{cont_orthog c funcs}
\l C_{\lambda}(x),C_{\lambda'}(x)\r=|W_\lambda|\cdot|F|\cdot\delta_{\lambda\lambda'},
\\\label{cont_orthog s funcs}
\l S_{\lambda}(x),S_{\lambda'}(x)\r=|W|\cdot|F|\cdot\delta_{\lambda\lambda'},
\end{gather}
where $\delta_{\lambda\lambda'}$ is the Kronecker delta, $|W|$ is the order of Weyl group, $|W_{\lambda}|$ is the size of Weyl group orbit, and $|F|$ is the volume of fundamental regions (formulas for $|F|$ are found in~\cite{HrivnakPatera2009}).

The functions of the families $C$ and $S$ are complete on the fundamental domain
(completeness follows from the completeness of the exponential functions).
%More precisely, there does not exist a function $\phi(x)$ such that
%{$\langle\phi(x),\phi(x)\rangle>0$} and at the same time $\langle\phi(x),\psi(x)\rangle=0$ for all functions $\psi(x)$ from the same family.

%%%%%%%%%%%%%%%%%%%%%%%%%%%%%%%%%%%%%%%%%%%%%%%%%%%%%%%%%%%%%%%
\section{Multivariate orthogonal polynomials corresponding to orbit functions}\label{sec_substitutions}
%%%%%%%%%%%%%%%%%%%%%%%%%%%%%%%%%%%%%%%%%%%%%%%%%%%%%%%%%%%%%%%
In this section we describe three substitutions of variables
\begin{gather*}
\mathfrak{T} :\  \R^n\ \longrightarrow\   \C^n;\qquad
(x_1,\dots,x_n)\ \mapsto\ (X_1,\dots,X_n),
\end{gather*}
that transform the $C(x)$- and $S(x)$-orbit functions into their polynomial forms in $n$ variables.

It directly follows from the orthogonality of the orbit functions~(\ref{cont_orthog c funcs}) and (\ref{cont_orthog s funcs})
that such polynomials are orthogonal on the domain $\tilde F$, where $\tilde F$ is the image of the fundamental region $F$ under the transformation $\mathfrak{T}$.
The corresponding integration weight function can be found using the Jacobian ${\rm det}^{-1}\left(\frac{D(X)}{D(x)}\right)$ of the transformation~$\mathfrak{T}$.

%%%%%%%%%%%%%%%%%%%%%%%%%%%%%%%%%%%%%%%%%%%%%%%%%%%%%%%%%%%%%%%
\subsection{Polynomials obtained by exponential substitution}
%%%%%%%%%%%%%%%%%%%%%%%%%%%%%%%%%%%%%%%%%%%%%%%%%%%%%%
The first substitution of variables is rather straightforward
\begin{gather}\label{subst_exp}
X_j:=e^{2\pi i x_j}, \qquad x_j\in\R,\quad j=1,2,\dots,n.
\end{gather}
Polynomial summands are products $\prod\limits^n_{j=1}X_j^{\mu_j}$, where $\mu_j\in\Z$ are components of the orbit points relative to a suitable basis. Under this transformation orbit function, $C_\lambda(x)$ and $S_\lambda(x)$, given by~(\ref{Cdef}) and (\ref{Sdef}), become Laurent polynomials in $n$ variables $X_j$, where $j=\{1,2,\dots,n\}$.

The exponential substitution polynomials are complex-valued in general, admit negative powers, and have all their coefficients equal to one in $C$-polynomials, and $1$ or $-1$ in $S$-polynomials.

Common trigonometric identities can be viewed as identities between $C$- and $S$-orbit functions of one variable. It is likely that identities between $C$- and $S$-orbit functions of more than one variable could also be found.

%It was shown in~\cite{Koornwinder1-4} that $C$- and $S$-polynomials of two variables can be found from completely different considerations.
%In~Example~\ref{example_koorwinder}, we demonstrated that the polynomials of~\cite{Koornwinder1-4} coincide with polynomials obtained by exponential substitution from orbit functions of~$A_2$.

%%%%%%%%%%%%%%%%%%%%%%%%%%%%%%%%%%%%%%%%%%%%%%%%%%%%%%%%%%%%%%%
\subsection{Polynomials obtained by trigonometric substitution}
%%%%%%%%%%%%%%%%%%%%%%%%%%%%%%%%%%%%%%%%%%%%%%%%%%%%%%%%%%%%%%%
The $W$-orbits of many simple Lie groups (but not all!) have an additional property that admits a~truly trigonometric substitution of variables. In order that $W_\lambda(L)$ is such an orbit we have to have the following,
\begin{gather}\label{bothweights}
\pm\mu\in W_\lambda(L)\qquad   \text{ for all }\ \mu\in W_\lambda(L).
\end{gather}
The pair of corresponding terms of the function of $W_\lambda(L)$ can be combined:
\begin{gather*}
e^{2\pi i\langle\mu,x\rangle}+e^{-2\pi i\langle\mu,x\rangle}
         =2\cos(2\pi\langle\mu,x\rangle)\in C_\lambda(x),\\
e^{2\pi i\langle\mu,x\rangle}-e^{-2\pi i\langle\mu,x\rangle}
         =2i\sin(2\pi\langle\mu,x\rangle)\in S_\lambda(x),
\end{gather*}
so that $C_\lambda(x)$ and $S_\lambda(x)$ become linear combinations of cosines and sines. Using common trigonometric identities, these cosines and sines can be expressed through the lowest ones, which are thus chosen as the new polynomial variables.

Obviously the important question is when \eqref{bothweights} is valid.
In Lie theory, the answer is known for representations, and extends without reservation to $W$-orbits.
The property that assures validity of \eqref{bothweights} is self-contragrediency of the representation of the highest weight $\lambda$ (see for example \cite{McP}).

All $W$-orbits of the following Lie groups have the property \eqref{bothweights}, hence admit the trigonometric substitutions:
\begin{gather}\label{list}
A_1,\quad B_n\ (n\geq3),\quad C_n\ (n\geq2),\quad D_{2n}\ (n\geq2),\quad E_7,\quad E_8,\quad F_4,\quad G_2.
\end{gather}
Note that in the case of $A_1$, this is precisely the trigonometric substitution made for Chebyshev polynomials of the first and second kind.

The representations/orbits of the remaining Lie groups that also have the property \eqref{bothweights} are listed in \cite{McP}.

%%%%%%%%%%%%%%%%%%%%%%%%%%%%%%%%%%%%%%%%%%%%%%%%%%%%%%%%%%%%%%%
\subsection{Polynomials obtained by recursive method}
%%%%%%%%%%%%%%%%%%%%%%%%%%%%%%%%%%%%%%%%%%%%%%%%%%%%%%
In the previous subsection, the substitution of variables transforming orbit functions into polynomials can be seen as generalization of the trigonometric variables of classical Chebyshev polynomials. Here, the substitution of variables is also a generalization of the trigonometric variables of classical Chebyshev polynomials, but of a different kind: It works uniformly for simple Lie algebras of all types, not only those of~\eqref{list}.

The fundamental weights of a simple Lie algebra $L_n$ of rank $n$ are the basis vectors $\omega_k$, $k=1,2,\dots,n$, of the $\omega$-basis of $\R^n$. We choose $C$-functions of the $n$ fundamental weights as the $n$ new variables:
\begin{gather}\label{variables}
X_j:=X_j(x):=C_{\omega_j}(x),\qquad j=1,2,\dots,n,\quad x\in\R^n;
\end{gather}
completed by one more variable, the lowest $S$-function,
\begin{gather}
\mathbf{S}:=S_\rho(x),\qquad x\in\R^n,\qquad \rho=(1,1,\dots,1)=\sum_{m=1}^n\omega_m,
\end{gather}
which cannot be constructed by multiplying the other variables.

The recursive construction of $C$-\emph{polynomials} begins by multiplying the variables $X_j$ and $C$-functions and decomposing their products into sums of $C$-polynomials. A judicious choice of the sequence of products allows one to find ever higher degree $C$-polynomials.

First, \textit{generic recursion relations} are found as the decomposition of products $X_{j}C_{(a_1,a_2,\dots,a_n)}$ with `sufficiently large' $a_1,a_2,\dots,a_n$ (i.e.\ all $C$-functions in the decomposition should correspond to generic points).
Then the rest of necessary recursions (`\textit{additional}') are constructed.
An~efficient way to find the decompositions is to work with products of Weyl group orbits, rather than with orbit functions. Their decomposition has been studied, and many examples have been described in \cite{HLP}.
Note that these recursion relations are always linear and the corresponding matrix is triangular.
The procedure is exemplified below for $L=A_1,\ A_2,\ A_3,\ C_2,$ and $G_2$. Also generic recursion relations are shown for $B_3$ and $C_3$.

Results of the recursive procedures can be summarized as follows.

\begin{proposition}\label{prop_c_polyn}
Any irreducible  $C$-function and any character $\chi_\lambda$ of a simple Lie group $G$
can be represented as a polynomial of $C$-functions of the fundamental weights $\omega_1,\dots,\omega_n$,
i.e.\ a~polynomial in the variables $X_1, X_2,\dots, X_n$.
\end{proposition}

\begin{proof}
Let us fix the dominance order (natural partial order) on the weight lattice:
\begin{gather*}
\lambda\succcurlyeq\mu\quad \Leftrightarrow \quad (\lambda-\mu)\in Q^+,
\end{gather*}
where $Q^+=\sum\limits_{\alpha\in \Delta^+}\N\alpha$, $\Delta^+$-positive roots.

Then we say that
\begin{gather*}
X_1^{\lambda_1}X_2^{\lambda_2}\cdots X_n^{\lambda_n} \succcurlyeq X_1^{\mu_1}X_2^{\mu_2}\cdots X_n^{\mu_n}.
\end{gather*}

As soon as the above ordering is fixed we can order our polynomials $C_{\lambda}(X)$, $\lambda=\lambda_1\w_1+\dots+\lambda_n\w_n$.
Then the above proposition follows from the method of induction with respect to weight lattice point $\lambda$.

The basis of induction is formed by the additional recursions, i.e.,
by all such expansions of the products $X_jC_{\lambda}$ that contain at least one orbit function $C_{\tilde{\lambda}}$,
where $\tilde{\lambda}$ have at least one zero coordinate in $\w$-basis.

As far as $\lambda$ is given in $\omega$-basis and $x$ is given in $\check{\alpha}$-basis
the expansion of the product $X_j(x)C_{\lambda}(x)$ is equivalent to the expansion of the Weyl orbits product $W_{\w_j}W_{\lambda}$.
Here exists unique $\lambda'\in W_\lambda$ such that $\forall \mu\in W_\lambda$  we have $\mu\preccurlyeq\lambda'$.
Then the general form of the recurrence relation is following
\begin{gather}\label{ind_step}
X_jC_{\lambda}=C_{(\lambda'_1,\lambda'_2,\dots\lambda'_j+1,\dots,\lambda'_n)}+L(C_{\preccurlyeq \lambda'}),\qquad j=1,2,\dots,n.
\end{gather}
Where $L(C_{\preccurlyeq \lambda'})$ is $\Z$-linear combination of orbit functions $C_\mu$, with $\mu\preccurlyeq\lambda'$.
The formula~\eqref{ind_step} implies the induction step for each $j=1,2,\dots,n$.

Therefore we proved that $C_\lambda$ is the polynomial of $C$-functions of the fundamental weights $\omega_1,\dots,\omega_n$.
The same statement for $\chi_\lambda$ follows from the fact that character is the linear combination of certain $C$-functions.
\end{proof}

The recursive construction of $S$-\emph{polynomials} starts by multiplying the variables $\mathbf{S}$ and $X_j$ and decomposing their products into sums of $S$-polynomials. Again a judicious choice of the sequence of products allows one to find ever higher degree $S$-polynomials. The procedure is exemplified below for $L=A_2,\ C_2,$ and $G_2$.
However, let us point out that, the higher the rank of the underlying Lie algebra, the recursive procedure for $S$-polynomials becomes more laborious, in comparison with the similar procedure for $C$-polynomials.
This is caused by the presence of negative terms in $S$-polynomials, and by the fact that all $S$-functions have the maximal number of exponential summands.
In calculating the decompositions of products of polynomials there are frequent cancelations of terms.

Fortunately, there is an alternative to the recursive procedure for $S$-polynomials. Once the $C$-polynomials have been calculated, they can be used in~\eqref{thecharacter} for finding $S$-polynomials as sums of $C$-polynomials multiplied by the variable $\mathbf{S}$.
\begin{remark}
It is important to note that in practice, polynomials $\frac{S_{\lambda}}{\mathbf{S}}$ should be used instead of~$S_{\lambda}$.
This is caused by the fact that the polynomial $S_{\lambda}$ depends on $n+1$ variables $\mathbf{S},X_1,\dots,X_n$ but it is defined for $n$-dimensional Euclidean space.
The fraction $\frac{S_{\lambda}}{\mathbf{S}}$ allows us to avoid this confusion.
The function $\mathbf{S}(X)$ has no zeros inside $F$ and at the boundary of $F$ both $\mathbf{S}$ and $S_{\lambda}$  have zero simultaneously,
so the fraction $\frac{S_{\lambda}}{\mathbf{S}}$ is well defined. In fact this means that we are working with the character instead of $S$-orbit function.
\end{remark}

\begin{remark}
There are two easy and practical checks on recursion relations applicable to all simple Lie algebras.
The first one is the equality of numbers of exponential terms in $S$- or $C$-functions on both sides of a recursion relation
(the numbers of exponential terms are calculated using the sizes of Weyl group orbits).

The second check is the equality of congruence numbers $\#$.
The character $\chi_\lambda$, the representation $\lambda$,
and the $C_\lambda$- and $S_\lambda$-polynomials, as well as all the points in one Weyl group orbit,
can be assigned to a congruence class specified by the congruence number $\#(\lambda)$,
which is the number (label) of the coset containing $\lambda$ in the lattice $P$ with respect to the root lattice.
In other words, our congruence numbers $\#$ are the elements of the quotient group $P/Q$.
The number of possible distinct values of $\#$ is equal to the order of the center of $G$.
The congruence numbers add up during multiplication.
In particular, all the $C$-polynomials in \eqref{thecharacter} must be from the same congruence class.
More generally, all the terms in the decomposition of a product of the polynomials belong to the same class.
This criterion is trivial for three of the simple Lie groups: $G_2$, $F_4$ and $E_8$,
since they have only one congruence class.

The number of exponential terms in a character $\chi_\lambda(x)$ is equal to the dimension of the representation with the highest weight $\lambda$.
Hence the number of terms in each of the $C$-polynomials in~\eqref{thecharacter} have to add up to the dimension of the representation.
%This is used in the case of $S$-polynomials in the numerator of \eqref{thecharacter}.
%In the case of $C$-polynomials, one also needs to know the number of terms in each $C$-function involved.
%There is an easy general rule \cite{HrivnakPatera2009}. For each of the specific Lie algebras of rank considered in the article, we list the sizes of the $W$-orbits.
\end{remark}
%%%%%%%%%%%%%%%%%%%%%%%%%%%%%%%%%%%%%%%%%%%%%%%%%%%%%%%%%%%%%%%%%%%%%%%%%%%%%%%%%%%%%%%%%%%%%%%%%%%
\begin{table}

\centering
\begin{tabular}{|c|c|c|c|c|c|c|c|}
\cline{6-8}
\multicolumn{5}{c|}{\quad}                                  &$\lambda$             &$A_3$ & $B_3$ and $C_3$\\
\cline{6-8}
\multicolumn{5}{c|}{\quad}                                  &$(\star,\star,\star)$ &24    &48\\
\cline{1-4}\cline{6-8}
$\lambda$             & $A_2$ & $C_2$ & $G_2$  &\qquad\qquad&$(\star,\star,0)$     &12    &24\\
\cline{1-4}\cline{6-8}
$(\star,\star)$       &   6   &   8   &   12               &&$(\star,0,\star)$     &12    &24\\
\cline{1-4}\cline{6-8}
$(\star,0)$           &   3   &   4   &    6               &&$(0,\star,\star)$     &12    &24\\
\cline{1-4}\cline{6-8}
$(0,\star)$           &   3   &   4   &    6               &&$(\star,0,0)$         &4     &6\\
\cline{1-4}\cline{6-8}
\multicolumn{5}{c|}{\quad}                                  &$(0,\star,0)$         &6     &12\\
\cline{6-8}
\multicolumn{5}{c|}{\quad}                                  &$(0,0,\star)$         &4     &8\\
\cline{6-8}
\end{tabular}
\caption{Number of points in $W$-orbits for the specific Lie algebras considered in this paper.
The orbits are labeled by their dominant weights $\lambda$, where the symbol  `$\star$'  stands for any positive integer.}\label{orbitsizes}
\end{table}
%%%%%%%%%%%%%%%%%%%%%%%%%%%%%%%%%%%%%%%%%%%%%%%%%%%%%%%%%%%%%%%%%%%%%%%%%%%%%%%%%%%%%%%%%%%%%%%%%%%

%%%%%%%%%%%%%%%%%%%%%%%%%%%%%%%%%%%%%%%%%%%%%%%%%%%%%%%%%%%%%%%
\section{Relations between orbit function polynomials and Macdonald polynomials}\label{sec_Macdonald}
%%%%%%%%%%%%%%%%%%%%%%%%%%%%%%%%%%%%%%%%%%%%%%%%%%%%%%%%%%%%%%%
In this section we establish connections between orbit functions
and known multivariate orthogonal polynomials.
In particular it will be shown that $C$-polynomials play the role of building blocks
in the theory of symmetric and antisymmetric polynomials%~\checkit{\cite{HeckO, Heck, Mac1, Mac2, Mac3, KVl}}
(Schur polynomials, Jacobi polynomials of many variables, Macdonald symmetric polynomials).

\subsection{Monomial symmetric polynomials}

Orbit functions $C_\lambda$ are a certain modification of \textit{monomial symmetric polynomials}
\begin{gather}\label{mon}
\mathfrak{m}_\lambda(y)=\sum_{\mu\in W_\lambda} y^\mu \equiv \sum_{\mu\in W_\lambda} y_1^{\mu_1}y_2^{\mu_2}\cdots y_n^{\mu_n},\qquad \lambda\in P^+.
\end{gather}
This relation follows from the substitution~\eqref{subst_exp} introduced in the previous section
(here $y_j=X_j$, $j=1,2,\dots,n$).

For studying symmetric orthogonal polynomials one usually replaces $y^\mu$ by $e^\mu$
($e^\mu$ is considered as a function on Euclidean space
$e^\mu(x)=e^{\langle \mu,x \rangle} =e^{\mu_1x_1+\cdots +\mu_nx_n}$).
Then we obtain modified $C$-orbit functions also called \textit{monomial symmetric functions}
\begin{gather}\label{mod}
\hat{\mathfrak{m}}_\lambda(x)=\sum_{\mu\in W_\lambda} e^{\langle \mu,x\rangle} = \sum_{\mu\in W_\lambda}e^{\mu_1x_1+\cdots +\mu_nx_n}.
\end{gather}
The functions $\hat{\mathfrak{m}}_\lambda(x)$ are also eigenfunctions of the Laplace operator.
Note that if we take integral orthogonal coordinates %($\alpha_j=e_j-e_{j+1}\in\R^{n+1}$)
$m_1,m_2,\ldots, m_n$ in the $A_n$ case in such a~way that $m_1\ge m_2\ge \cdots\ge m_n\ge 0$,
then Laurent polynomials $\mathfrak{m}_\lambda(y)$ and $\hat{\mathfrak{m}}_\lambda (x)$ turn into usual (not
Laurent) polynomials.

Jacobi polynomials in one variable are well-known orthogonal
polynomials of the theory of special functions and multivariate Jacobi polynomials are symmetric Laurent polynomials,
which are defined by means of $\hat{\mathfrak{m}}_\lambda$, $\lambda\in P_+$, see~\cite{HeckO, Heck}.

\subsection{Macdonald symmetric polynomials}
We consider Macdonald symmetric (Laurent) polynomials,
which are a quantum analogue of Jacobi polynomialsa and constructed by means of monomial symmetric polynomials.

Let $A$ denote the group algebra over $\R$ of the free Abelian group $P$,
then Weyl group $W$ acts on $P$ and $A$.
Further let $A^W$ denote the subalgebra of $W$-invariant elements of $A$
and monomial symmetric functions $\hat{\mathfrak{m}}_\lambda$ defined in previous section
form a basis of $A^W$.

We introduce a real number $q$, $0\leqslant q < 1$ and with every root $\alpha$ from the root system~$R$ of~$G$
we associate a variable $t_\alpha$ such that $t_\alpha=t_{w\alpha}$, $w\in W$.

Let
${\mathbb C}_{q,t_\alpha}$ be the field of rational functions in $q$ and $t_\alpha$
and let ${\mathbb C}_{q,t_\alpha}[e^{x_1},\ldots ,e^{x_n}]$ denote the set of Laurent polynomials in functions from the previous section
$e^{x_1},e^{x_2},\ldots ,e^{x_n}$ with coefficients from ${\mathbb C}(q,t_\alpha)$.

The constant term $a_0$ of a function $\sum\limits_\lambda a_\lambda e^\lambda$ we denote by $[\sum\limits_\lambda a_\lambda e^\lambda]_0$.

For $f,g\in{\mathbb C}_{q,t_\alpha}[e^{x_1},\ldots ,e^{x_n}]$ we determine an inner product{\samepage
\begin{gather*}\label{scal}
\langle f,g \rangle_{q,p_\alpha}=|W|^{-1}[f{\overline{g}} \mathbf{\Delta}]_0,
\quad \text{where}\quad
{\mathbf\Delta}=\prod_{\alpha\in R} \prod_{i=1}^\infty \frac{1-q^{2i}e^\alpha}{1-t_\alpha^2q^{2i}e^\alpha},
\end{gather*}
and the bar over $g$ denotes the linear involution determined by $\overline{e^\lambda}=e^{-\lambda}$.}

The Macdonald polynomials $P_\lambda$, $\lambda\in P^+$ are uniquely defined by the following theorem proved by I.~Macdonald~\cite{Mac3}.
\begin{theorem}\label{theorem_macdonald} There exists a unique family $P_\lambda \in {\mathbb C}_{q,t_\alpha}[e^{x_1},e^{x_2},\ldots ,e^{x_n}]^W$,
$\lambda\in P_+$, satisfying the conditions
\begin{gather*}
1)\ P_\lambda=\hat{\mathfrak{m}}_\lambda +\sum_{\mu<\lambda} a_\lambda^\mu \hat{\mathfrak{m}}_\mu ,\qquad  a_\lambda^\mu\in {\mathbb C}_{q,t_\alpha},\\
2)\ \langle P_\lambda,P_\mu \rangle_{q,t_\alpha}=0,\qquad  {\rm if}\quad \lambda\ne\mu .
\end{gather*}
\end{theorem}
In other words the Macdonald polynomials are obtained via orthogonalization of the basis for $A^W$.
The orthogonality property of the Macdonald polynomials is proved by showing that the Macdonald polynomials are eigenvectors
for an algebra of commuting self adjoint operators with one-dimensional eigenspaces,
and using the fact that eigenspaces for different eigenvalues must be orthogonal.

Replacing $e^{x_j}$ by $y_j$, $j=1,2,\ldots ,n$, in $P_\lambda$ we
obtain (for each fixed values of $q$ and $t_\alpha$) orthogonal
symmetric polynomials which are called {\it Macdonald symmetric
polynomials}.

Replacing $y_j$ by $e^{2\pi{\rm i}x_j}$, $j=1,2,\ldots ,n$, in
Macdonald polynomials we obtain orthogonal functions which are
finite linear combinations of $C$-orbit functions.

Some special values of $q$ and $t_\alpha$ reduce Macdonald polynomials to $C$- and $S$- orbit functions:
\begin{itemize}
\item
If $t_\alpha=1$, then independently of $q$ we have
$P_\lambda(y)=\hat{\mathfrak{m}}_\lambda(y)=C_\lambda(x)|_{y=2\pi i x}$.

\item
If $\forall\alpha\in R$\;  $t_\alpha=q$, then
$P_\lambda(y)=\chi_\lambda(y)=\frac{S_\lambda(x)}{S_\rho(x)}\big|_{y=x}$.
\end{itemize}

\begin{corollary}
Polynomial forms of $C$- an $S$-functions introduced in Section~{\rm \ref{sec_substitutions}}
by substitution~\eqref{subst_exp}
are partial cases of the Macdonald symmetric polynomials.
Polynomial forms of orbit functions obtained by the trigonometric substitution and
substitution~\eqref{variables} are equivalent to these Macdonald symmetric polynomials.
\end{corollary}

Theorem~\ref{theorem_macdonald} and Proposition~\ref{prop_c_polyn} imply the following conclusion.

\begin{corollary}
Macdonald symmetric polynomials can be represented as polynomials
in fundamental $C$-functions: $X_1=C_{\w_1},\; \dots,\; X_n=C_{\w_n}$.
Such an approach produce \textit{not Laurent} polynomials, but polynomials with all nonnegative powers.
\end{corollary}

\begin{remark}
All $C$- and $S$-orthogonal polynomials described in Section~\ref{sec_substitutions}
inherit from orbit functions important discretization properties.
An uniform discretization of these polynomials follows from their invariance with respect
to the affine Weyl group of $G$ and from the well-established discretization of the fundamental region $F(G)$,
see~\cite{HrivnakPatera2009} for details.

It is worth to mention that behind each polynomial we have special functions (orbit functions).
It makes our orthogonal polynomials richer and gives a number of advantages, one of them is cubature formula
introduced in~\cite{MoodyPatera2010}.
\end{remark}

Due to the relation between orbit functions and Macdonald polynomials
the mentioned discretization method can be adopted and applied to Macdonald polynomials.

The orbit functions do not require any sophisticated theory to work with.
In general, for a given group $G$, most of the properties of orbit functions come from
the Weyl group and the irreducible character of $G$ or from the basic notions of the representation theory.
This fact makes multivariate polynomials obtained from orbit functions attractive and handy for a wide range of applications,
such as discrete and interpolation problems, cubature formulas~\cite{MoodyPatera2010}, reduction of polynomials~\cite{NPST}, etc.

For the application reason below we present recurrence relations and lowest polynomials for the simple Lie groups
$A_1$, $A_2$, $A_3$, $C_2$, $G_2$, $C_3$, $B_3$.
All, but two last cases, contain both generic and additional recursions.
For groups $C_3$ and $B_3$ we obtain only generic recurrence relations
and lowest polynomials necessary to solve them are available in~\cite{NPST}.

The content of the following sections is also motivated by
the fact that calculation of additional recurrences is not suitable for complete computer automatization.
%This caused by the property that the Weyl group orbit of a point $\lambda$
%essentially depends on if $\lambda$ has any zero coordinates in $\w$-basis.
However, as soon as additional recurrences were obtained,
all other calculations concerning polynomials and their applications
become very algorithmic and easily can be done by computer algebra packages for Lie theory.
In following sections it is also shown in all details how the recursive algorithm proposed in Section~\ref{sec_substitutions} works.

%%%%%%%%%%%%%%%%%%%%%%%%%%%%%%%%%%%%%%%%%%%%%%%%%%%%%%
\section{Discussion}
%%%%%%%%%%%%%%%%%%%%%%%%%%%%%%%%%%%%%%%%%%%%%%%%%%%%%%%%%%%%%%%%%%%%%
\begin{itemize}\itemsep=-1pt
%\begin{enumerate}
\item
Recent years Lie groups became a backbone of a segment of the theory of orthogonal polynomials of many variables
and monomial symmetric functions/orbit functions became building blocks of multivariate orthogonal polynomials.
In particular, this idea is illustrated by orthogonal polynomials of orbit functions,
that are the special cases of the Macdonald polynomials~\cite{Mac3},
which in turn are the special cases of more general notion of Koornwinder polynomials~\cite{Koornwinder1993}.

%\item
%Some of the properties of orbit functions translate readily into
%properties of Chebyshev polynomials of many variables. However there
%are other properties whose discovery from the theory of
%polynomials is difficult to imagine. As an example, consider the
%decomposition of the Chebyshev polynomial of the second kind (characters) into
%the sum of Chebyshev polynomials of the first kind (C-polynomials). In one variable,
%it is a familiar problem that can be solved by elementary
%means. For two and more variables, the problem turns out to be
%equivalent to a more general question about representations of
%simple Lie groups \cite{MP82}.

\item
There is an alternative way to our construction of the polynomials in all but in the $A_n$ cases. The crucial substitution \eqref{variables} can be replaced by
\begin{equation}\label{var-chi}
X_k:=\chi_{\w_k}(x),\qquad k=1,2,\dots,n\,.
\end{equation}
In \eqref{var-chi} the variables are characters of irreducible representations with highest weights given as the fundamental weights, while  in \eqref{variables} the variables are $C$-functions of the fundamental weights. Only for $A_n$ the two coincide, \mbox{$C_{\w_k}(x)=\chi_{\w_k}(x)$} for all $k=1,\dots,n$ and for all $x\in\R^n$.
Already for the rank two cases other than $A_2$ there is a difference. Indeed, \eqref{var-chi} reads as follows,
\begin{gather*}
\begin{array}{@{}lll}
C_2 :& X_1 = \chi_{\omega_1}(x) = C_{\omega_1}(x),                            & X_2 = \chi_{\omega_2}(x) = C_{\omega_2}(x) + 2;\\
G_2 :& X_1 = \chi_{\omega_1}(x) = C_{\omega_1}(x) + C_{\omega_2}(x) + 2,\quad & X_2 = \chi_{\omega_2}(x) = C_{\omega_2}(x) + 1.
\end{array}
\end{gather*}
Since products of characters decompose into their sum, the recursive construction can proceed, but the polynomials will be different. Both approaches clearly can rightfully claim to be generalizaiions of classical Chebyshev polynomials.

\item
Our approach to the derivation of multidimensional orthogonal
polynomials hinges on the knowledge of appropriate recursion
relations. The basic mathematical property underlying the existence
of the recursion relation is the complete decomposability of
products of the orbit functions. Numerous examples of the
decompositions of products of orbit functions, involving also other
Lie groups than ${\rm SU}(n)$, were shown elsewhere
\cite{KlimykPatera2006,KlimykPatera2007-1}. An equivalent problem is
the decomposition of products of Weyl group orbits \cite{HLP}.
\smallskip

\item
The possibility to discretize the polynomials is a consequence of the known
discretization of orbit functions.
For orbit functions it is a simpler problem, in that it is carried out in the real Euclidean space $\R^n$.
In principle, it carries over to the polynomials.
But variables of the polynomials happen to be on the maximal torus of the underlying Lie group.
Only in the cases of \eqref{list}, the variables are real (the imaginary unit multiplying the $S$-functions can be normalized away).
For $A_n$ with $n>1$, only some of the orbit functions of the congruence class 0  are real valued.
Practical aspects of discretization deserve to be further investigated.
\smallskip

\item
For simplicity of formulation, we insisted throughout this paper
that the underlying Lie group be simple. The extension to compact
semisimple Lie groups and their Lie algebras is straightforward. Thus,
orbit functions are products of orbit functions of simple
constituents, and different types of orbit functions can be mixed.
\smallskip

\item
Polynomials formed from $E$-functions by the same substitution of
variables should be equally interesting once $n>1$. We know of no
analogs of such polynomials in the standard theory of polynomials
with more than one variable. Intuitively, they would be formed as
sums $C+S$ polynomials, Their domain of orthogonality is twice as large as that of Chebyshev polynomials. The $E$-functions have been studied in~\cite{HrivnakPatera2009, KlimykPatera2008, Patera}.

%%%%%%%%%% new remark %%%%%%%%%%%%%%%%%%
\item
The generating functions for characters of irreducible finite  dimensional
representations of simple Lie groups were invented in~\cite{PS}. They contain  wealth of
information about the characters  and therefore also about the orbit functions and hence
about the  polynomials. In principle they are not limited by the number of  variables.
Unfortunately, these functions grow rapidly in complexity  for higher groups. The example
of the generating function of $G_2$  characters is rather convincing \cite{GS}. However
there is an aspect  to the character generators that is not matched in other methods of
constructing the polynomials. They provide rather specific information  about existence
of syzygies (identities) in the polynomial rings.
\end{itemize}

\begin{center}
APPENDIX
\end{center}

\appendix
%%%%%%%%%%%%%%%%%%%%%%%%%%%%%%%%%%%%%%%%%%%%%%%%%%%%%%
%%%%%%%%%%%%%%%%%%%%%%%%%%%%%%%%%%%%%%%%%%%%%%%%%%%%%%
%%%%%%%%%%%%%%%%%%%%%%%%%%%%%%%%%%%%%%%%%%%%%%%%%%%%%%

%%%%%%%%%%%%%%%%%%%%%%%%%%%%%%%%%%%%%%%%%%%%%%%%%%%%%%%%%%%%%%%
\section{Orbit functions of $A_1$, their polynomial forms and Chebyshev polynomials}\label{sec_A_1}
%%%%%%%%%%%%%%%%%%%%%%%%%%%%%%%%%%%%%%%%%%%%%%%%%%%%%%%%%%%%%%%
A number of multivariate generalizations of classical Chebyshev polynomials are available in literature~\cite{DunnLidl1980,LY, Lidl, Rivlin1974, suetin},
the aim of this section is to show in all details how Chebyshev polynomials appear as particular case of the multivariate polynomials proposed in this paper.
First we recall that well-known classical Chebyshev polynomials
can be  obtained independently using only the properties of $C$- and $S$-orbit functions of the Lie group $A_1$,
see~\cite{NPT} for detailes.
%As a consequence, one-to-one correspondence to classical polynomials becomes evident.
The $C$-polynomials yielded by our approach are naturally normalized in a different way than the classical polynomials.
It makes the correspondence between $C$- and $S$-polynomials a direct special case of the general properties \eqref{thecharacter} and \eqref{thecharacterinv}.

We start a derivation of $A_1$ polynomials from the beginning in a way that emphasizes the underlying Lie algebra $A_1$ and, more importantly, in a way that directly generalizes to simple Lie algebras of any rank $n$ and any type, resulting in polynomials of $n$ variables and of a new type for each Lie algebra.

The construction yields a different normalization of polynomials (form of Dickson polynomials) and their trigonometric variables than is common for classical Chebyshev polynomials. In~$1D$, no new polynomials emerge than those equivalent to the classical Chebyshev polynomials of the first and second kind. Our insight into the structure of the problem for the general simple Lie group was gained in this construction. We are inclined to refer to the Chebyshev polynomials derived in this paper as the canonical ones.

The underlying Lie algebra $A_1$ is often denoted $su(2)$.  In fact, this case is so simple that the presence of the Lie algebras never needed to be acknowledged\footnote{The Lie algebra $su(2)$ is the backbone of the angular momentum theory in quantum physics. The normalization of its generator implied here is common in mathematics. In particular, $m$ equals to twice the angular momentum in physics.}.

The size of an orbit of $A_1$ with the highest weight $(m)$ in $\omega$-basis is either 2, if $m>0$, or 1, if $m=0$. A weight belongs to the congruence class specified by the value $\#(m)=m\mod2$.  In particular, all exponents of monomials in a polynomial $C_m$ or $S_m$ have the parity of $m$. The dimension of the representation $(m)$ is $d_{(m)}=m+1$.

The orbit functions of $A_1$ are of two types:
\begin{gather}
C_m(x)= e^{2\pi i m x}+e^{-2\pi i m x}=2\cos(2\pi m x),\qquad
     x\in \R,\quad m\in\Z^{> 0};           \label{c-func_A1}\\
S_m(x)= e^{2\pi i m x}-e^{-2\pi i m x}=2i\sin(2\pi m x),\qquad
     x\in \R,\quad m\in\Z^{> 0}.             \label{s-func_A1}
\end{gather}
When $m=0$, the general definitions~\eqref{Cdef} and~\eqref{Sdef} give $C_0(x)=1$, and $S_0(x)=0$ for all $x$,
but~\eqref{c-func_A1} gives $C_0(x)=2$. We keep this convention in this section.

The simplest substitution of variables that would transform the orbit functions into a polynomial is $Z^m=e^{2\pi i m x}$. Exponents in such a polynomial are the integers $m$ and $-m$. Chebyshev polynomials can be represented in symmetric way: with symmetric positive and negative powers, and with all coefficients equal to $1$ or $-1$. Instead, we introduce new variables $X$ and $S$ as follows:
\begin{equation}
\begin{gathered}
X:=C_1(x)= e^{2\pi i x}+e^{-2\pi i x}=2\cos(2\pi x),\\
S:=S_1(x)= e^{2\pi i x}-e^{-2\pi i x}=2i\sin(2\pi x).
\end{gathered}
\end{equation}
Polynomials can now be constructed recursively in the degrees of $X$ and $S$ by calculating the decompositions of products of appropriate orbit functions.
`Generic' recursion relations are those where one of the first degree polynomials, $X$ or $S$, multiplies the generic polynomial~$C_m$ or~$S_m$, i.e. $m\geq1$. Omitting the dependence on $x$ from the symbols, we have the generic recursion relations
\begin{gather*}
\begin{array}{@{}l@{\qquad}l}
XC_m=C_{m+1}+C_{m-1},&    SC_m=S_{m+1}-S_{m-1},\\
XS_m=S_{m+1}+S_{m-1},&    SS_m=C_{m+1}-C_{m-1},
\end{array}
\qquad m\geqslant 1.
\end{gather*}

When solving recursion relations for $C$-polynomials, we need to start from the lowest ones:
\begin{gather*}
\begin{array}{@{}llll}
X^2 =C_2+C_0=C_2+2    &\Longrightarrow & C_2=X^2-2,&\nonumber\\
XC_2=C_3+C_1=C_3+X    &\Longrightarrow & C_3=XC_2-X=X^3-3X,&\nonumber\\
XC_m=C_{m+1}+C_{m-1}  &\Longrightarrow & C_{m+1}=XC_m-C_{m-1},&m\geqslant 3.\nonumber
\end{array}
\end{gather*}
Several lowest results are in Table~\ref{tableA1}.
Hence we conclude that $C_m=2T_m$, for $m=0,1,\dots$.

There are also recursion relations for $C$-polynomials resulting from products of two $S$-po\-ly\-no\-mials:
\begin{gather*}
\begin{array}{@{}llll}
S^2 =C_2-2            &\Longrightarrow & C_2=S^2+2,           &\nonumber\\
%SS_2=C_3-X            &\Longrightarrow & C_3 =S^3+3X,         &\nonumber\\
SS_m=C_{m+1}-C_{m-1}  &\Longrightarrow & C_{m+1}=SS_m+C_{m-1},& m\geqslant2.\nonumber
\end{array}
\end{gather*}

Note that each $C_m$ can be written in two ways, as a polynomial of degree $m$ in $X$, and as a~polynomial of the same degree involving $S$ and $X$. Equating the two expressions for~$C_m$, we obtain a trigonometric identity for each $m$. For example, from the two ways of writing $C_2$, we find
{\samepage
\begin{gather*}
X^2-S^2=4\quad\Longleftrightarrow\quad \sin^2(2\pi x)+\cos^2(2\pi x)=1.
\end{gather*}
Our $S$ is defined to be purely imaginary, hence the negative sign at $S^2$.}

Analogous relations yield polynomial expressions for $S_m$,
\begin{gather*}
\begin{array}{@{}llll}
XS=S_2                &\Longrightarrow &  S_2=XS&\\
XS_2=S_3+S            &\Longrightarrow &  S_3=XS_2-S=X^2S-S&\\
XS_m=S_{m+1}+S_{m-1}  &\Longrightarrow &  S_{m+1}=SS_m-S_{m-1},  & m\geqslant2;\\[1ex]
SC_2=S_3-S            &\Longrightarrow &  S_3=SC_2+S=S^3+3S&\\
SC_m=S_{m+1}+S_{m-1}  &\Longrightarrow &  S_{m+1}=SS_m-C_{m-1},  & m\geqslant2.
\end{array}
\end{gather*}

A fundamental relation between $S$- and $C$-polynomials (appropriately normalized) is the special case of~\eqref{thecharacter}.
It eliminates the need to find $S$-polynomials recursively, provided the $C$-polynomials have been found.
The character $\chi_m(x)$ of an irreducible representation of $A_1$ of dimension $m+1$ is known explicitly for all $m\geqslant0$.
There are two ways to write the character: as the ratio of $S$-functions, and as the sum of $C$-functions. Explicitly, that is
\begin{gather}\label{A1character}
\chi_m(x) = \frac{S_{m + 1}(x)}{S(x)} = C_m(x) + C_{m - 2}(x) + \cdots +
    \begin{cases}
    C_2(x) + C_0\;   &\text {if $m$ even},\\
    C_3(x) + X(x)\; &\text {if $m$ odd}.\\
    \end{cases}
\end{gather}
Note that~\eqref{A1character} is the Chebyshev polynomial of the second kind
$U_m(x)$.

\begin{remark}
Note that in the character formula we used $C_0=1$, while for $C$-polynomials we used $C_0=2$.
It is just a question of normalization of orbit function $C_0$.
Here we used it in order to obtain classical form of the Chebyshev polynomials.
More generally for any simple $G$, it is sometime convenient to scale up orbit functions of non-generic point,
say $\lambda$, by the factor equal to the order of the stabilizer of $\lambda$ in the Weyl group $W$.
\end{remark}

\begin{remark}
The main argument in favor of our normalization of Chebyshev polynomials is that
polynomials $C_m$ from Table~\ref{tableA1} are Dickson polynomials
(it is well known that them are equivalent to Chebyshev polynomials over the complex numbers).
It is easy to prove~(see e.g.~\cite{NPT}) that Weyl group of~$A_n$ is equivalent to~$S_{n+1}$,
therefore it is natural to consider multivariate $C$-polynomials of $A_n$ as n-dimensional generalizations of
Dickson polynomials (as permutation polynomials).
Also our form of Dickson--Chebyshev polynomials makes them
the lowest special case of~\eqref{thecharacter} and~\eqref{thecharacterinv} without additional adjustments
and it appears more `natural' because, for example, the equality $C_2^2=C_4+2$ would not hold for $T_2$ and $T_4$.
\end{remark}

%%%%%%%%%%%%%%%%%%%%%%%%%%%%%%%%%%%%%%%
\begin{table}
\centering
\begin{tabular}{|l|l|c|l|l|}
\cline{1-2}\cline{4-5}
\multicolumn{2}{|c|}{$C$-polynomials} &\hspace{1cm} & \multicolumn{2}{c|}{$S$-polynomials$\phantom{\sum^{A^1}}$}\\
\cline{1-2}\cline{4-5}
%$C_{0}$ & $2$                     && $S_{0}$ & 0\\
%\cline{1-2}\cline{4-5}
\multicolumn{2}{|l|}{$\#=0$} &\hspace{1cm} & \multicolumn{2}{l|}{$\#=0$$\phantom{\sum^{A^1}}$}\\
\cline{1-2}\cline{4-5}
$C_{2}$ & $X^2-2$                  && $S_{1}$ & $1$$\phantom{\sum^{A^1}}$\\
\cline{1-2}\cline{4-5}
$C_{4}$ & $X^4-4X^2+2$             && $S_{3}$ & $X^2-1$$\phantom{\sum^{A^1}}$\\
\cline{1-2}\cline{4-5}
$C_{6}$ & $X^6-6X^4+9X^2-2$        && $S_{5}$ & $X^4-3X^2+1$$\phantom{\sum^{A^1}}$\\
\cline{1-2}\cline{4-5}
$C_{8}$ & $X^8-8X^6+20X^4-16X^2+2$$\phantom{\sum^{A^1}}$ && $S_{7}$ & $X^6-5X^4+6X^2-1$\\
\cline{1-2}\cline{4-5}
\multicolumn{2}{|l|}{$\#=1$} &\hspace{1cm} & \multicolumn{2}{l|}{$\#=1$$\phantom{\sum^{A^1}}$}\\
\cline{1-2}\cline{4-5}
$C_{1}$ & $X$                      && $S_{2}$ & $X$$\phantom{\sum^{A^1}}$\\
\cline{1-2}\cline{4-5}
$C_{3}$ & $X^3-3X$                 && $S_{4}$ & $X^3-2X$$\phantom{\sum^{A^1}}$\\
\cline{1-2}\cline{4-5}
$C_{5}$ & $X^5-5X^3+5X$            && $S_{6}$ & $X^5-4X^3+3X$$\phantom{\sum^{A^1}}$\\
\cline{1-2}\cline{4-5}
$C_{7}$ & $X^7-7X^5+14X^3-7X$      && $S_{8}$ & $X^7-6X^5+10X^3-4X$$\phantom{\sum^{A^1}}$\\
\cline{1-2}\cline{4-5}
\end{tabular}
\caption{The irreducible $C$- and $S$-polynomials of $A_1$ of degrees up to 8.} \label{tableA1}
\end{table}
%%%%%%%%%%%%%%%%%%%%%%%%%%%%%%%%%%%%%%%

%%%%%%%%%%%%%%%%%%%%%%%%%%%%%%%%%%%%%%%%%%%%%%%%%%%%%%%%%%%%%%%%%%%%%%
\section{Recursion relations for $A_2$ orbit functions and polynomials}\label{sec_A_2}
%%%%%%%%%%%%%%%%%%%%%%%%%%%%%%%%%%%%%%%%%%%%%%%%%%%%%%%%%%%%%%%%%%%%%%
In the previous section, variables $X$ and $S$ played an almost symmetrical role.
This is not the case when the rank of the Lie algebra $L_n$ exceeds 1.
The number of variables of type $X$ is equal to $n$,
the number of exponential functions comprising such variables is a (small) divisor of the order $|W|$ of the Weyl group.
In contrast, the variable $S$ is unique for all $L_n$.
It is a sum of the maximal number of exponential functions, namely $|W|$.

The~variables of the $A_2$ polynomials are the $C$-functions of the lowest dominant weights $\w_1=(1,0)$, $\w_2=(0,1)$, and the unique lowest $S$-function whose dominant weight is $(1,1)$.  The~variables are denoted as follows,
\begin{gather}\label{variablesA2}
X_1:=C_{(1,0)}(x_1,x_2),\qquad X_2:=C_{(0,1)}(x_1,x_2),\qquad S:=S_{(1,1)}(x_1,x_2).
\end{gather}
We omit writing $(x_1,x_2)$ at the symbols of orbit functions for simplicity of notations.

In addition to the obvious polynomials $X_1$, $X_2$, $X_1^2$, $X_1X_2$, and $X_2^2$, we recursively find the rest of the  $A_2$-polynomials. The degree of the polynomial $C_{(a,b)}$ equals $a+b$. The degree of $S_{(a,b)}$ is also $a+b$ provided $ab\neq 0$, otherwise the $S$-polynomials are zero.

Due to the $A_2$ outer automorphism, polynomials $C_{(a,b)}$ and $C_{(b,a)}$ are related by the interchange of variables
$X_1\leftrightarrow X_2$ (i.e. $C_{(a,b)}(X_1,X_2)=C_{(b,a)}(X_2,X_1)$).

In general, each term in an irreducible polynomial, equivalently each weight of an orbit, must belong to the same congruence class specified by the congruence number $\#$. For $A_2$-weight $(a,b)$, we have
\begin{gather}\label{cogruenceA2}
\#(a,b)=(a+2b)\mod3.
\end{gather}
Hence, irreducible orbit functions have a well defined value of $\#$. For $A_2$-orbit functions, we have
$\#(C_{(a,b)})=\#(S_{(a,b)})=(a+2b)\mod3$. Consequently, there are three classes of polynomials corresponding to $\#=0,1,2$. During multiplication, the congruence numbers add up $\mod3$. A~product of irreducible orbits decomposes into the sum of orbits belonging to the same congruence class.

The sizes of the irreducible orbits of $W(A_2)$ are found in Table~\ref{orbitsizes}. The dimension $d_{(a,b)}$ of the representation of $A_2$ with the highest weight $(a,b)$ is given by
\begin{gather}\label{dimensionA2}
d_{(a,b)}=\tfrac12(a+1)(b+1)(a+b+2).
\end{gather}

%%%%%%%%%%%%%%%%%%%%%%%%%%%%%%%%%%%%%%%%%%%%%%%%%%%%%%%%%%%%%%%%%%%%%%
\subsection{Recursion relations for $\boldsymbol{C}$-function polynomials of $\boldsymbol{A_2}$}
There are two 4-term generic recursion relations for $C$-functions. They are obtained as the decomposition of the products of $X$ and $Y$, each being a sum of three exponential functions, with a generic $C$-function which is the sum of $|W(A_2)|=6$ exponential terms. We call  $C_{(a,b)}$ generic, provided it is a sum of 6 distinct exponential terms. In order that a recursion relation be generic, the product has to decompose into three distinct generic (i.e. 6-term) $C$-functions
\begin{gather*}\label{A2generic}
X_1C_{(a,b)}=C_{(a+1,b)}+C_{(a-1,b+1)}+C_{(a,b-1)},\qquad a,b\geqslant2;\\
X_2C_{(a,b)}=C_{(a,b+1)}+C_{(a+1,b-1)}+C_{(a-1,b)},\qquad a,b\geqslant2.
\end{gather*}
Before generic recursion relations can be used, the special recursion relations
for particular values $a,b\in\{0,1\}$ need to be solved recursively starting from the lowest ones:
\begin{alignat*}{3}
& {X_1^2  = C_{(2,0)} + 2X_2, \quad X_2^2  = C_{(0,2)} + 2X_1},\qquad &&
X_1X_2  = C_{(1,1)} + 3;&
\\
& X_1C_{(1,1)}  = C_{(2,1)} + 2C_{(0,2)} + 2X_1,\qquad &&
X_2C_{(1,1)}  = C_{(1,2)} + 2C_{(2,0)} + 2X_2;&
\\[1ex]
& \text{for }a\geqslant2:&&& \\
& X_1C_{(a,1)} = C_{(a + 1,1)} + C_{(a - 1,2)} + 2C_{(a,0)},\qquad &&
X_2C_{(a,1)} = C_{(a,2)} + 2C_{(a + 1,0)} + C_{(a - 1,1)};&
\\
& X_1C_{(a,0)} = C_{(a + 1,0)} + C_{(a - 1,1)},\qquad &&
X_2C_{(a,0)}  = C_{(a,1)} + C_{(a - 1,0)};&
\\[1ex]
& \text{for }b\geqslant2: &&&
\\
& X_1C_{(1,b)}  = C_{(2,b)} + 2C_{(0,b + 1)} + C_{(1,b - 1)},\qquad &&
X_2C_{(1,b)}  = C_{(1,b + 1)} + C_{(2,b - 1)} + 2C_{(0,b)};&
\\
& X_1C_{(0,b)}  = C_{(1,b)} + C_{(0,b - 1)},\qquad &&
X_2C_{(0,b)}  = C_{(0,b + 1)} + C_{(1,b - 1)}.&
\end{alignat*}
Using the symmetry of orbit functions with respect to the permutation of the components of dominant weights,
we obtain analogous polynomials $C_{(a,0)}$ and $C_{(a,1)}$ for all $a\in\N$.
Then the 4-term special recursion relations are solved yielding $C_{(2,b)}$ and $C_{(a,2)}$ for all $a,b\in\N$.
After that, the generic recursion relations should be used.

%%%%%%%%%%%%%%%%%%%%%%%%%%%%%%%%%%%%%%%%%%%%%%%%%%%%%%%%%%%%%%%%%%%%%%
\subsection{Recursion relations for $\boldsymbol{S}$-function polynomials of $\boldsymbol{A_2}$}
%%%%%%%%%%%%%%%%%%%%%%%%%%%%%%%%%%%%%%%%%%%%%%%%%%%%%%
Polynomials $S_{(a,0)}=S_{(0,b)}=0$ vanish for all $a$ and $b$.
There are two 4-term generic recursion relations for $S$-functions,
\begin{gather*}%\label{A2Sgeneric}
X_1S_{(a,b)}=S_{(a+1,b)}+S_{(a-1,b+1)}+S_{(a,b-1)},\qquad a,b\geqslant2;\\
X_2S_{(a,b)}=S_{(a,b+1)}+S_{(a+1,b-1)}+S_{(a-1,b)},\qquad a,b\geqslant2.
\end{gather*}

Then there are the special 3- and 4-term recursion relations for particular values of $a=1$ and/or $b=1$:
\begin{gather*}
\begin{array}{@{}lll}
X_1S=S_{(2,1)},                       & X_2S=S_{(1,2)};\\
X_1S_{(1,2)}=S_{(2,2)}+S,             & X_2S_{(1,2)}=S_{(1,3)}+S_{(2,1)};\\
X_1S_{(2,1)}=S_{(3,1)}+S_{(1,2)},     & X_2S_{(2,1)}=S_{(2,2)}+S;\\
X_1S_{(3,1)}=S_{(4,1)}+S_{(2,2)},\qquad   & X_2S_{(3,1)}=S_{(2,1)}+S_{(3,2)}.
\end{array}
\end{gather*}
Further on, generic recursion relations can be used.

There are additional recursion relations for $C$-polynomials, even if not necessarily used for finding the polynomials:
\begin{gather*}
SS=C_{(2,2)}-2C_{(0,3)}-2C_{(3,0)}+2C_{(1,1)}-6,\\
SS_{(2,1)}=C_{(3,2)}-C_{(1,3)}-2C_{(4,0)}+2C_{(0,2)}-2C_{(1,0)}+C_{(2,1)},\\
SS_{(1,2)}=C_{(2,3)}-C_{(3,1)}-2C_{(0,4)}+2C_{(2,0)}-2C_{(0,1)}+C_{(1,2)},\\
SS_{(2,2)}=C_{(3,3)}-C_{(4,1)}- C_{(1,4)}+2C_{(0,3)}+2C_{(3,0)}-C_{(1,1)},
\\[1ex]
\text{for }a,b\geqslant3:\\
SS_{(a,b)} = C_{(a + 1,b + 1)} - C_{(a + 2,b - 1)} - C_{(a - 1,b + 2)} + C_{(a - 2,b + 1)} + C_{(a + 1,b - 2)} - C_{(a - 1,b - 1)}.
\end{gather*}
These are interesting for a different reason. Indeed, a particular $S_{(a,b)}$ can be written as a~linear combination of $C$-functions {\it divided} by~$S$.
From~\eqref{thecharacter}, we find that $S_{(a,b)}$ is written as a~polynomial of $C$-functions {\it multiplied} by~$S$.

%%%%%%%%%%%%%%%%%%%%%%%%%%%%%%%%%%%%%%%%%%%%%%%%%%%%%%%%%%%%%%%%%%%%%%
\subsection{The character of $\boldsymbol{A_2}$}
%%%%%%%%%%%%%%%%%%%%%%%%%%%%%%%%%%%%%%%%%%%%%%%%%%%%%%
The character $\chi_{(a,b)}$ is given either as a fraction of $S$-functions (Weyl character formula)
or as a~linear combination of the $C$-function.
In the $A_2$ case the general formula \eqref{thecharacter} is specialized
\begin{gather*}\label{characterA2}
\chi_{(a,b)}(x,y)=\frac{S_{(a+1,b+1)}(x,y)}{S_{(1,1)}(x,y)}=C_{(a,b)}(x,y)+\sum_\lambda m_\lambda C_\lambda(x,y).
\end{gather*}
The summation extends over the dominant weights that have positive multiplicities $m_\lambda$
in the case of $\chi_{(a,b)}$. The coefficients (dominant weight multiplicities) are tabulated in \cite{BMP}
for the 50 first $\chi_{(a,b)}$ in each congruence class of $A_2$.
The first few characters for the congruence class~$\#=0$~are:
\begin{gather*}
\chi_{(0,0)}=C_{(0,0)}=1,\\
\chi_{(1,1)}=C_{(1,1)}+2C_{(0,0)},\\
\chi_{(3,0)}=C_{(3,0)}+C_{(1,1)}+C_{(0,0)},\\
\chi_{(0,3)}=C_{(0,3)}+C_{(1,1)}+C_{(0,0)},\\
\chi_{(2,2)}=C_{(2,2)}+C_{(0,3)}+C_{(3,0)}+2C_{(1,1)}+3C_{(0,0)},\\
\chi_{(1,4)}=C_{(1,4)}+C_{(2,2))}+2C_{(0,3)}+C_{(3,0)}+2C_{(1,1)}+2C_{(0,0)},\\
\chi_{(3,3)}=C_{(3,3)} + C_{(4,1)} + C_{(1,4)} + 2C_{(2,2))} + 2C_{(0,3)} + 2C_{(3,0)} + 3C_{(1,1)} + 4C_{(0,0)},\\
\chi_{(6,0)}=C_{(6,0)}+C_{(4,1)}+C_{(2,2))}+C_{(0,3)}+C_{(3,0)}+C_{(1,1)}+C_{(0,0)}.
\end{gather*}
The equalities must satisfy two relatively simple conditions:
(i) The dominant weights on both sides must have the same congruence number \eqref{cogruenceA2}, and
(ii) the number of exponential terms in a character $\chi_{(a,b)}$ is known to be the dimension
\eqref{dimensionA2} of the irreducible representation $(a,b)$.
Therefore, the sizes of the orbit functions on the right side have to add up to the dimension.
For $\#=1$:
\begin{gather*}
\chi_{(1,0)}=C_{(1,0)},\\
\chi_{(0,2)}=C_{(0,2)}+C_{(1,0)},\\
\chi_{(2,1)}=C_{(2,1)}+C_{(0,2)}+2C_{(1,0)},\\
\chi_{(1,3)}=C_{(1,3)}+C_{(2,1)}+2C_{(0,2)}+2C_{(1,0)},\\
\chi_{(4,0)}=C_{(4,0)}+C_{(2,1)}+C_{(0,2)}+C_{(1,0)},\\
\chi_{(0,5)}=C_{(0,5)}+C_{(1,3)}+C_{(2,1)}+C_{(0,2)}+C_{(1,0)}.
\end{gather*}

For $\#=2$, it suffices to interchange the component of all dominant weights in the equalities for $\#=1$.
Thus no independent calculation is needed.
%%%%%%%%%%%%%%%%%%%%%%%%%%%%%%%%%%%%%%%
\begin{table}
\centering
\begin{tabular}{|l|l|c|l|l|}
\cline{1-2}\cline{4-5}
\multicolumn{2}{|c|}{$C$-polynomials} &\hspace{1cm} & \multicolumn{2}{c|}{$S$-polynomials$\phantom{\sum^{A^1}}$}\\
\cline{1-2}\cline{4-5}
\multicolumn{2}{|l|}{$\#=0$} &\hspace{1cm} & \multicolumn{2}{l|}{$\#=0$$\phantom{\sum^{A^1}}$}\\
\cline{1-2}\cline{4-5}
$C_{(1,1)}$ & $X_1X_2 - 3$                                 && $S_{(2,2)}$ & $X_1X_2-1$$\phantom{\sum^{A^1}}$\\
\cline{1-2}\cline{4-5}
$C_{(3,0)}$ & $X_1^3 - 3X_1X_2 + 3$                        && $S_{(1,4)}$ & $X_2^3 - 2X_1X_2 + 1$$\phantom{\sum^{A^1}}$\\
\cline{1-2}\cline{4-5}
$C_{(2,2)}$ & $X_1^2X_2^2 - 2X_1^3 - 2X_2^3 + 4X_1X_2 - 3$ && $S_{(3,3)}$ & $X_1^2X_2^2 - X_1^3 - X_2^3$$\phantom{\sum^{A^1}}$\\
\cline{1-2}\cline{4-5}
\multicolumn{2}{|l|}{$\#=1$} &\hspace{1cm} & \multicolumn{2}{l|}{$\#=1$$\phantom{\sum^{A^1}}$}\\
\cline{1-2}\cline{4-5}
%$C_{(0,0)}$ & $1$                                          && $S_{(1,1)}$ & $S$\\
%\cline{1-2}\cline{4-5}
$C_{(1,0)}$ & $X_1$                                        && $S_{(2,1)}$ & $X_1$$\phantom{\sum^{A^1}}$\\
\cline{1-2}\cline{4-5}
%$C_{(2,0)}$ & $X_1^2 - 2X_2$                               && $S_{(1,3)}$ & $X_2^2-X_1$\\
%\cline{1-2}\cline{4-5}
$C_{(0,2)}$ & $X_2^2 - 2X_1$                               && $S_{(1,3)}$ & $X_2^2-X_1$$\phantom{\sum^{A^1}}$\\
\cline{1-2}\cline{4-5}
$C_{(2,1)}$ & $X_1^2X_2 - 2X_2^2 - X_1$                    && $S_{(3,2)}$ & $X_1^2X_2 - X_2^2 - X_1$$\phantom{\sum^{A^1}}$\\
\cline{1-2}\cline{4-5}
$C_{(1,3)}$ & $X_1X_2^3 - 3X_1^2X_2 - X_2^2 + 5X_1$        && $S_{(2,4)}$ & $X_1X_2^3 - 2X_1^2X_2 - X_2^2 + 2X_2$$\phantom{\sum^{A^1}}$\\
\cline{1-2}\cline{4-5}
$C_{(4,0)}$ & $X_1^4 - 4X_1^2X_2 + 2X_2^2 + 4X_1$          && $S_{(5,1)}$ & $X_1^4 - 3X_1^2X_2 + X_2^2 + X_1 + X_2$$\phantom{\sum^{A^1}}$\!\!\!\!\!\!\!\!\!\\
\cline{1-2}\cline{4-5}
\end{tabular}
\caption{The irreducible $C$- and $S$-polynomials of $A_2$ of degree up to 4.
From any polynomial $C_{(a,b)}$ or $S_{(a,b)}$ we obtain $C_{(b,a)}$ or $S_{(b,a)}$ respectively
by interchanging $X_1$ and $X_2$.} \label{tableA2}
\end{table}
%%%%%%%%%%%%%%%%%%%%%%%%%%%%%%%%%%%%%%%

\begin{example}\label{example_koorwinder}
Comparison of polynomials obtained by our exponential substitution method in the case of $A_2$, with the results of~\cite{Koornwinder1-4}(III), reveals coincidence of the polynomials in both cases, as demonstrated in this example.
Moreover, the polynomials considered here are the partial case of the Mecdonald polynomials.

%Consider the  orbits $W_{(0,m)}$, $W_{(m,0)}$, and the orbit of the generic point $W_{(m_1,m_2)}$, where $m,m_1,m_2\in\Z^{>0}$.
%\begin{gather*}
%\begin{array}{l}
%W_{(0,m)}(A_2)=\{(0, m), (  - m, 0), (m,  - m)\},
%\\[1ex]
%W_{(m,0)}(A_2)=\{(m, 0), ( - m, m), (0,  - m)\},
%\\[1ex]
%W_{(m_1,m_2)}(A_2)=\{
%(m_1, m_2)^+,\ ( - m_1, m_1 + m_2)^-,\ (m_1 + m_2,  - m_2)^-,
%\\[1ex]
%\phantom{W_{(m_1,m_2)}(A_2)=\{}
%( - m_2,  - m_1)^-,\ (  - m_1 - m_2, m_1)^+,\ (m_2,  - m_1 - m_2)^+\}.
%\end{array}
%\end{gather*}

Suppose $x=(x_1,x_2)$ is given in the $\alpha$-basis, then $C$-functions $A_2$ of assume the form:
\begin{gather}
C_{(0,0)}(x)=1,
\nonumber\\
C_{(0,m)}(x)=\overline{C_{(m,0)}(x)}=e^{ - 2\pi i mx_1} + e^{2\pi i mx_1}e^{ - 2\pi i mx_2} + e^{2\pi i mx_2},
\nonumber\\
C_{(m_1,m_2)}(x)=e^{2\pi i m_1x_1}e^{2\pi i m_2x_2} + e^{ - 2\pi i m_1x_1}e^{2\pi i (m_1 + m_2)x_2} +{}
\nonumber\\
\phantom{C_{(m_1,m_2)}(x)=} e^{2\pi i (m_1 + m_2)x_1}e^{ - 2\pi i m_2x_2} + e^{ - 2\pi i m_2x_1}e^{ - 2\pi i m_1x_2}+{}
\nonumber\\
\phantom{C_{(m_1,m_2)}(x)=}e^{ - 2\pi i (m_1 + m_2)x_1}e^{2\pi i m_1x_2} + e^{2\pi i m_2x_1}e^{ - 2\pi i (m_1 + m_2)x_2},
\nonumber\\
S_{(m_1,m_2)}(x)=e^{2\pi i m_1x_1}e^{2\pi i m_2x_2} - e^{ - 2\pi i m_1x_1}e^{2\pi i (m_1 + m_2)x_2}-{}
\nonumber\\
\phantom{S_{(m_1,m_2)}(x)=}e^{2\pi i (m_1 + m_2)x_1}e^{ - 2\pi i m_2x_2} - e^{ - 2\pi i m_2x_1}e^{ - 2\pi i m_1x_2}+{}
\nonumber\\
\phantom{S_{(m_1,m_2)}(x)=}e^{ - 2\pi i (m_1 + m_2)x_1}e^{2\pi i m_1x_2} + e^{2\pi i m_2x_1}e^{ - 2\pi i (m_1 + m_2)x_2}. \label{C_S_for_A2}
\end{gather}
%Using the exponential substitution~(\ref{subst_exp}), we have the following corresponding polynomials
%\begin{gather*}
%\begin{array}{l}
%C_{(0,0)}=1,\\[1ex]
%C_{0,m}=\overline{C_{0,m}}=X_1^{ - m} + X_1^{m}X_2^{ - m} + X_2^{m},
%\\[1ex]
%C_{(m_1,m_2)}
%=X_1^{m_1}X_2^{m_2}+X_1^{-m_1}X_2^{(m_1+m_2)}+X_1^{(m_1+m_2)}X_2^{-m_2}+
%\\[1ex]
%\phantom{C_{(m_1,m_2)}=}
%X_1^{-m_1}X_2^{-m_2}+X_1^{-(m_1+m_2)}X_2^{m_1}+X_1^{m_2}X_2^{-(m_1+m_2)},
%\\[1ex]
%S_{(m_1,m_2)}
%=X_1^{m_1}X_2^{m_2}-X_1^{-m_1}X_2^{(m_1+m_2)}-X_1^{(m_1+m_2)}X_2^{-m_2}-
%\\[1ex]
%\phantom{S_{(m_1,m_2)}=}
%X_1^{-m_1}X_2^{-m_2}+X_1^{-(m_1+m_2)}X_2^{m_1}+X_1^{m_2}X_2^{-(m_1+m_2)}.
%\end{array}
%\end{gather*}

The polynomials $e^+$ and $e^-$ given in (2.6) of~\cite[III]{Koornwinder1-4} coincide with those
in~(\ref{C_S_for_A2}) whenever the correspondence $\sigma=2\pi x_1$, $\tau=2\pi x_2$ is set up.
Thus both our orbit function polynomials of~$A_2$ and $e^{\pm}$ of~\cite[III]{Koornwinder1-4} are orthogonal on the interior of Steiner's hypocycloid.

It is noteworthy that the regular tessellation of the plane by
equilateral triangles considered in~\cite{Koornwinder1-4} is the
standard tiling of the weight lattice of $A_2$. The fundamental
region~$R$ of~\cite{Koornwinder1-4} coincides with the fundamental
region $F(A_2)$ in our notations. The corresponding isometry group
is the affine Weyl group of~$A_2$.

Furthermore, continuing the comparison with~\cite{Koornwinder1-4}, we should emphasize that it is known that orbit functions are eigenfunctions not only of the Laplace operator but also of the differential operators built from the elementary symmetric polynomials, see~\cite{KlimykPatera2006, KlimykPatera2007-1}.
\end{example}

%%%%%%%%%%%%%%%%%%%%%%%%%%%%%%%%%%%%%%%%%%%%%%%%%%%%%%%%%%%%%%%%%%%%%
\section{Recursion relations for $C_2$ orbit functions}\label{sec_C_2}
%%%%%%%%%%%%%%%%%%%%%%%%%%%%%%%%%%%%%%%%%%%%%%%%%%%%%%%%%%%%%%%%%%%%%%
There are two congruence classes of $C_2$ orbit functions/polynomials. For $C_2$ weight $(a,b)$ (dominant or not), we have
\begin{gather}
\#(a,b)=a\mod2
\end{gather}
The~dimension $d_{(a,b)}$ of an irreducible representation of $C_2$ with the highest weight $(a,b)$ is given~by
\begin{gather}\label{dimensionG2}
d_{(a,b)}=\tfrac16(a+1)(b+1)(2a+b+3)(a+b+2).
\end{gather}

In multiplying the polynomials, congruence numbers add up $\mod2$.
Character in the case of $C_2$ is given by~\eqref{thecharacter},
where the $C$- and $S$-functions are those of $C_2$,
as are the coefficients $m_\lambda$ (also tabulated in \cite{BMP}).

%\samepage{
We denote the variables of the $C_2$-polynomials by
\begin{gather*}
X_1:=C_{(1,0)}(x,y),\qquad X_2:=C_{(0,1)}(x,y),\qquad \text{and} \qquad S:=S_{(1,1)}(x,y),
\end{gather*}
often omitting $(x,y)$ from the symbols. The variable $S$ cannot be built out of $X_1$ and $X_2$.
%}
Although the variables are denoted by the same symbols as in the case of~$A_2$ (and also $G_2$ below),
they are very different. Thus $X_1$ and $X_2$ contain 4 exponential terms and $S$ contains 8 terms.
The~congruence number~$\#$ of $X_1$ and $S$ is 1, while that of $X_2$ is 0.

%%%%%%%%%%%%%%%%%%%%%%%%%%%%%%%%%%%%%%%%%%%%%%%%%%%%%%%%%%%%%%%%%
\subsection{Recursion relations for $\boldsymbol{C}$-functions of $\boldsymbol{C_2}$}
%%%%%%%%%%%%%%%%%%%%%%%%%%%%%%%%%%%%%%%%%%%%%%%%%%%%%%
The two generic recursion relations for $C$-functions of $C_2$ are
\begin{gather*}\label{C2generic}
X_1C_{(a,b)}=C_{(a+1,b)}+C_{(a-1,b+1)}+C_{(a+1,b-1)}+C_{(a-1,b)}, \qquad a,\ b\geqslant2;\\
X_2C_{(a,b)}=C_{(a,b+1)}+C_{(a+2,b-1)}+C_{(a-2,b+1)}+C_{(a,b-1)}, \qquad a\geqslant3,\ b\geqslant2.
\end{gather*}
The special recursion relations for $C$-functions involving low values of $a$ and $b$
have to be solved first starting from the lowest ones:
\begin{gather*}
\begin{array}{@{}ll}
X_1C_{(a,1)} =C_{(a+1,1)}+C_{(a-1,2)}+2C_{(a+1,0)}+C_{(a-1,1)},&a\geqslant2;\\
X_1C_{(a,0)} =C_{(a+1,0)}+C_{(a-1,1)} +C_{(a-1,0)},            &a\geqslant2;\\
X_1C_{(1,b)} =C_{(2,b)}+2C_{(0,b+1)}+C_{(2,b-1)}+2C_{(0,b)},   &b\geqslant2;\\
X_1C_{(0,b)} =C_{(1,b)}+C_{(1,b-1)},                           &b\geqslant2;\\
X_1C_{(1,1)} =C_{(2,1)}+2C_{(0,2)}+2C_{(2,0)}++2X_2 ;          &\\
\multicolumn{2}{@{}l}{
X_1X_2=C_{(1,1)}+2X_1;\qquad
X_1^2 =C_{(2,0)}+2X_2+4;}\\
%\end{array}
%\end{gather*}
%
%\begin{gather*}
%\begin{array}{@{}ll}
X_2C_{(a,1)} =C_{(a,2)}+2C_{(a+2,0)}+C_{(a-2,2)}+2C_{(a,0)},     &a\geqslant3;\\
X_2C_{(a,0)} =C_{(a,1)}+C_{(a-2,1)},                             &a\geqslant3;\\
X_2C_{(2,b)} =C_{(2,b+1)}+C_{(4,b-1)}+2C_{(0,b+1)}+C_{(2,b-1)},  &b\geqslant2;\\
X_2C_{(1,b)} =C_{(1,b+1)}+C_{(3,b-1)}+C_{(1,b-1)}+C_{(1,b)},     &b\geqslant2;\\
X_2C_{(0,b)} =C_{(0,b+1)}+C_{(2,b-1)}+C_{(0,b-1)},               &b\geqslant2;\\
X_2C_{(2,1)} =C_{(2,2)}+2C_{(4,0)}+C_{(0,4)}+2C_{(0,2)}+2C_{(2,0)};\quad &\\
X_2C_{(1,1)} =C_{(1,2)}+2C_{(3,0)}+C_{(1,1)}+2X_1;               &\\
\multicolumn{2}{@{}l}{
X_2C_{(2,0)} =C_{(2,1)}+2X_2;\qquad
X_2^2 =C_{(0,2)}+2C_{(2,0)}+4.}
\end{array}
\end{gather*}

The 3- and 4-term recursion relations are solved independently, giving us $C_{(0,b)}$, $C_{(a,0)}$, $C_{(1,b)}$,
and $C_{(a,1)}$ for all $a$ and $b$, e.g.~see Table~\ref{tableC2-Cpoly}.

%%%%%%%%%%%%%%%%%%%%%%%%%%%%%%%%%%%%%%%%%%%%%%
\begin{table}
\centering
\begin{tabular}{|l|l|}
\hline
\multicolumn{2}{|c|}{$C$-polynomials$\phantom{\sum^{A^1}}$}\\
\hline
\multicolumn{2}{|l|}{$\#=0$$\phantom{\sum^{A^1}}$}\\
\hline
$C_{(0,1)}$ & $X_2$$\phantom{\sum^{A^1}}$\\
\hline
$C_{(2,0)}$ & $X_1^2 - 2X_2 - 4$$\phantom{\sum^{A^1}}$\\
\hline
$C_{(2,1)}$ & $X_1^2X_2 - 2X_2^2 - 6X_2$$\phantom{\sum^{A^1}}$\\
\hline
$C_{(4,0)}$ & $4 - 4X_1^2 + X_1^4 + 8X_2 - 4X_1^2X_2 + 2X_2^2$$\phantom{\sum^{A^1}}$\\
\hline
$C_{(0,2)}$ & $4 - 2X_1^2 + 4X_2 + X_2^2$$\phantom{\sum^{A^1}}$\\
\hline
$C_{(0,3)}$ & $9X_2 - 3X_1^2X_2 + 6X_2^2 + X_2^3$$\phantom{\sum^{A^1}}$\\
\hline
$C_{(2,2)}$ & $ - 8 + 10X_1^2 - 2X_1^4 - 20X_2 + 8X_1^2X_2 - 12X_2^2 + X_1^2X_2^2 - 2X_2^3$$\phantom{\sum^{A^1}}$\\
\hline
$C_{(0,4)}$ & $4 - 8X_1^2 + 2X_1^4 + 16X_2 - 8X_1^2X_2 + 20X_2^2 - 4X_1^2X_2^2 + 8X_2^3 + X_2^4$$\phantom{\sum^{A^1}}$\\
\hline
\multicolumn{2}{|l|}{$\#=1$$\phantom{\sum^{A^1}}$}\\
\hline
$C_{(1,0)}$ & $X_1$$\phantom{\sum^{A^1}}$\\
\hline
$C_{(1,1)}$ & $X_1X_2 - 2X_1$$\phantom{\sum^{A^1}}$\\
\hline
$C_{(3,0)}$ & $X_1^3 - 3X_1X_2 - 3X_1$$\phantom{\sum^{A^1}}$\\
\hline
$C_{(3,1)}$ & $2X_1 - 4X_1X_2 + X_1^3X_2 - 3X_1X_2^2$$\phantom{\sum^{A^1}}$\\
\hline
$C_{(1,2)}$ & $6X_1 - 2X_1^3 + 3X_1X_2 + X_1X_2^2$$\phantom{\sum^{A^1}}$\\
\hline
$C_{(1,3)}$ & $ - 6X_1 + 2X_1^3 + 6X_1X_2 - 3X_1^3X_2 + 5X_1X_2^2 + X_1X_2^3$$\phantom{\sum^{A^1}}$\\
\hline
\end{tabular}
\caption{The irreducible $C$-polynomials of $C_2$ of degree up to 4.} \label{tableC2-Cpoly}
\end{table}
%%%%%%%%%%%%%%%%%%%%%%%%%%%%%%%%%%%%%%%%%%%%%%

%%%%%%%%%%%%%%%%%%%%%%%%%%%%%%%%%%%%%%%%%%%%%%%%%%%%%%%%%%%%%%%%%
\subsection{Recursion relations for $\boldsymbol{S}$-functions of $\boldsymbol{C_2}$}
%%%%%%%%%%%%%%%%%%%%%%%%%%%%%%%%%%%%%%%%%%%%%%%%%%%%%%
The generic relations for $S$-functions are readily obtained from those of $C$-functions by replacing~$C$ by $S$, and by making appropriate sign changes,
\begin{gather*}
X_1S_{(a,b)} =S_{(a+1,b)}-S_{(a-1,b+1)}+S_{(a+1,b-1)}-S_{(a-1,b)},   \qquad a,\ b\geqslant2;\\
X_2S_{(a,b)} =S_{(a,b+1)}-S_{(a+2,b-1)}+S_{(a-2,b+1)}-S_{(a,b-1)},   \qquad a\geqslant3,\ b\geqslant2.
\end{gather*}
The special recursion relations for $S$-functions involving low values of $a$ and $b$ have to be solved first, starting from the lowest ones:
\begin{gather*}
\begin{array}{@{}ll}
X_1S_{(a,1)}=S_{(a+1,1)}+S_{(a-1,2)}+S_{(a-1,1)},           & a\geqslant2;\\
X_2S_{(a,1)} =S_{(a,2)}+S_{(a-2,2)},                        & a\geqslant3;\\
%X_1S_{(2,b)} =S_{(3,b)}+S_{(1,b+1)}+S_{(3,b-1)}+S_{(1,b)},  & b\geqslant2;\\
X_2S_{(2,b)} =S_{(2,b+1)}+S_{(4,b-1)}+S_{(2,b-1)},          & b\geqslant2;\\
X_1S_{(1,b)} =S_{(2,b)}+S_{(2,b-1)},                        & b\geqslant2;\\
X_2S_{(1,b)} =S_{(1,b+1)}+S_{(3,b-1)}+S_{(1,b-1)}-S_{(1,b)},\;          & b\geqslant2;\\
%X_1S_{(2,1)} =S_{(3,1)}+S_{(1,2)}+S;                        \\
X_2S_{(2,1)} =S_{(2,2)};                                    \\
X_1S =S_{(2,1)};                                            \\
X_2S =S_{(1,2)}-S.
\\[1ex]
SS_{(a,b)}   =C_{(a + 1,b + 1)} - C_{(a - 1,b + 2)} - C_{(a + 3,b - 1)} - C_{(a - 3,b + 1)}\\
\phantom{SS_{(a,b)} =} + C_{(a - 1,b - 1)} - C_{(a + 1,b - 2)} + C_{(a + 3,b - 2)} + C_{(a - 3,b + 2)}, &a\geqslant4,\; b\geqslant3;\\
SS=C_{(2,2)} - 2C_{(0,3)} - 2C_{(4,0)} - 2C_{(2,0)} - 2X_2 + 2X_2 + 2C_{(2,1)} + 8.
\end{array}
\end{gather*}

All $C$-functions of $C_2$ are real-valued. Here are a few examples of $C_2$-characters:
\begin{gather*}
\#=0\colon\\
\chi_{(0,0)}=C_{(0,0)}=1;\\
\chi_{(0,1)}=1+C_{(0,1)}=1+X_2;\\
\chi_{(2,0)}=2+X_2+C_{(2,0)};\\
\chi_{(0,2)}=2+X_2+C_{(2,0)}+C_{(0,2)};\\
\chi_{(2,1)}=3+3X_2+2C_{(2,0)}+C_{(0,2)}+C_{(2,1)};\\
\chi_{(0,3)}=2+2X_2+C_{(2,0)}+C_{(0,2)}+C_{(2,1)}+C_{(0,3)};\\
\chi_{(4,0)}=3+2X_2+2C_{(2,0)}+C_{(0,2)}+C_{(2,1)}+C_{(4,0)};\\
\chi_{(2,2)}=5 + 4X_2 + 4C_{(2,0)} + 3C_{(0,2)} + 2C_{(2,1)} + C_{(0,3)} + C_{(4,0)} + C_{(2,2)}.\\[1ex]
%\end{gather*}
%\begin{gather*}
\#=1\colon\\
\chi_{(1,0)}=C_{(1,0)}=X_1;\\
\chi_{(1,1)}=2X_1+C_{(1,1)};\\
\chi_{(3,0)}=2X_1+C_{(1,1)}+C_{(3,0)};\\
\chi_{(1,2)}=3X_1+2C_{(1,1)}+C_{(3,0)}+C_{(1,2)};\\
\chi_{(3,1)}=4X_1+3C_{(1,1)}+2C_{(3,0)}+C_{(1,2)}+C_{(3,1)};\\
\chi_{(1,3)}=4X_1+3C_{(1,1)}+2C_{(3,0)}+2C_{(1,2)}+C_{(3,1)}+C_{(1,3)}.
\end{gather*}

Using these characters and Table~\ref{tableC2-Cpoly}, we can calculate all irreducible $S$-polynomials of degree up to four
with respect to the variables $X_1$ and $X_2$ using \eqref{thecharacterinv}.
Note that $\chi_{(0,4)}$ yields the polynomial of order five.

%%%%%%%%%%%%%%%%%%%%%%%%%%%%%%%%%%%%%%%%%%%%%%%%%%%
\begin{table}
\centering
\begin{tabular}{|l|l|}
\hline
\multicolumn{2}{|c|}{$S$-polynomials$\phantom{\sum^{A^1}}$}\\
\hline
\multicolumn{2}{|l|}{$\#=0$$\phantom{\sum^{A^1}}$}\\
\hline
$S_{(1,2)}$  &  $1+X_2$$\phantom{\sum^{A^1}}$\\
\hline
$S_{(3,1)}$  &  $-2+X_1^2-X_2$$\phantom{\sum^{A^1}}$\\
\hline
$S_{(1,3)}$  &  $2-X_1^2+3X_2+X_2^2$$\phantom{\sum^{A^1}}$\\
\hline
$S_{(3,2)}$  &  $-1-3X_2+X_1^2X_2-X_2^2$$\phantom{\sum^{A^1}}$\\
\hline
$S_{(1,4)}$  &  $2-X_1^2+7X_2-2X_1^2X_2+5X_2^2+X_2^3$$\phantom{\sum^{A^1}}$\\
\hline
$S_{(5,1)}$  &  $3-4X_1^2+X_1^4+4X_2-3X_1^2X_2+X_2^2$$\phantom{\sum^{A^1}}$\\
\hline
$S_{(3,3)}$  &  $-3 + 4X_1^2 - X_1^4 - 7X_2 + 3X_1^2X_2 - 5X_2^2 + X_1^2X_2^2 - X_2^3$$\phantom{\sum^{A^1}}$\\
\hline
\multicolumn{2}{|l|}{$\#=1$$\phantom{\sum^{A^1}}$}\\
\hline
$S_{(2,1)}$  &  $X_1$$\phantom{\sum^{A^1}}$\\
\hline
$S_{(2,2)}$  &  $X_1X_2$$\phantom{\sum^{A^1}}$\\
\hline
$S_{(4,1)}$  &  $ - 3X_1 + X_1^3 - 2X_1X_2$$\phantom{\sum^{A^1}}$\\
\hline
$S_{(2,3)}$  &  $2X_1 - X_1^3 + 2X_1X_2 + X_1X_2^2$$\phantom{\sum^{A^1}}$\\
\hline
$S_{(4,2)}$  &  $ - 4X_1X_2 + X_1^3X_2 - 2X_1X_2^2$$\phantom{\sum^{A^1}}$\\
\hline
$S_{(2,4)}$  &  $5X_1X_2 - 2X_1^3X_2 + 4X_1X_2^2 + X_1X_2^3$$\phantom{\sum^{A^1}}$\\
\hline
\end{tabular}
\caption{The irreducible $S$-polynomials of $C_2$ of degree up to 4.} \label{tableC2-Spoly}
\end{table}
%%%%%%%%%%%%%%%%%%%%%%%%%%%%%%%%%%%%%%%%%%%%%%%%%%%

%%%%%%%%%%%%%%%%%%%%%%%%%%%%%%%%%%%%%%%%%%%%%%%%%%%%%%%%%%%%%%%%%%%%%%
\section{Recursion relations for $G_2$ orbit functions}\label{sec_G_2}
%%%%%%%%%%%%%%%%%%%%%%%%%%%%%%%%%%%%%%%%%%%%%%%%%%%%%%%%%%%%%%%%%%%%%%
 All $G_2$ weights fall into the same congruence class $\#=0$. Thus there are no congruence classes to distinguish in $G_2$. The dimension of the irreducible representation  $(a,b)$ of $G_2$ is given by
\begin{gather}\label{dimG2}
d_{(a,b)}=\tfrac1{120}(a + 1)(b + 1)(a + b + 2)(2a + b + 3)(3a + b + 4)(3a + 2b + 5).
\end{gather}
The variables are the orbit functions of the two fundamental weights $\w_1=(1,0)$ and $\w_2=(0,1)$. Also $S(x,y):=S_{(1,1)}(x,y)$ serves as an independent variable{\sloppy
\begin{gather*}
X_1:=C_{(1,0)}(x,y),\qquad X_2:=C_{(0,1)}(x,y),\qquad S=S_{(1,1)}(x,y).
\end{gather*}
The variables $X_1$ and $X_2$ are the sums of 6 exponential terms, while $S$ has 12 terms, also  $C_{(0,0)}(x,y){=}1$.

}

%%%%%%%%%%%%%%%%%%%%%%%%%%%%%%%%%%%%%%%%%%%%%%%%%%%%%%%%
\subsection{Recursion relations for $\boldsymbol{C}$-functions of $\boldsymbol{G_2}$}
%%%%%%%%%%%%%%%%%%%%%%%%%%%%%%%%%%%%%%%%%%%%%%%%%%%%%%
There are two generic recursion relations for $C$-polynomials of $G_2$, each containing one product term and six $C$-polynomials:
\begin{gather*}
\text{for }a{\geqslant}3,\ b{\geqslant}4:
\\
X_1C_{(a,b)} = C_{(a + 1,b)} + C_{(a - 1,b + 3)} + C_{(a + 2,b - 3)} + C_{(a - 2,b + 3)} + C_{(a + 1,b - 3)} + C_{(a - 1,b)};
\\[1ex]
\text{for }a{\geqslant}2,\ b{\geqslant}3:
\\
X_2C_{(a,b)} = C_{(a,b + 1)} + C_{(a + 1,b - 1)} + C_{(a - 1,b + 2)} + C_{(a + 1,b - 2)} + C_{(a - 1,b + 1)} + C_{(a,b - 1)}.
\end{gather*}
%}

Specializing the first of the generic relations to either $a\in\{0,1,2\}$ or $b\in\{0,1,2,3\}$, we have
\begin{gather*}
X_1C_{(2,b)}=C_{(3,b)}+C_{(1,b+3)}+C_{(4,b-3)}+2C_{(0,b+3)}+C_{(3,b-3)} +C_{(1,b)};\\
X_1C_{(1,b)}=C_{(2,b)}+2C_{(0,b+3)}+C_{(3,b-3)}+C_{(2,b-3)}+C_{(1,b)} +2C_{(0,b)};\\
X_1C_{(0,b)}=C_{(1,b)}+C_{(2,b-3)}+C_{(1,b-3)};\\
X_1C_{(a,3)}=C_{(a+1,3)}+C_{(a-1,6)}+2C_{(a+2,0)}+C_{(a-2,6)}+2C_{(a+1,0)}+C_{(a-1,3)};\\
X_1C_{(a,2)}=C_{(a+1,2)}+C_{(a-1,5)}+C_{(a+1,1)}+C_{(a-2,5)}+C_{(a,1)}+C_{(a-1,2)};\\
X_1C_{(a,1)}=C_{(a+1,1)}+C_{(a-1,4)}+C_{(a-2,4)}+C_{(a-1,1)}+C_{(a,2)}+C_{(a-1,2)};\\
X_1C_{(a,0)}=C_{(a+1,0)}+C_{(a-1,3)}+C_{(a-2,3)}+C_{(a-1,0)};\\
%\end{array}
%\end{gather*}
%\begin{gather*}
%\begin{array}{l}
X_1C_{(2,3)}=C_{(3,3)}+C_{(1,6)}+2C_{(4,0)}+2C_{(0,6)}+2C_{(3,0)}+C_{(1,3)};\\
X_1C_{(2,2)}=C_{(3,2)}+C_{(1,5)}+C_{(0,5)}+C_{(1,2)}+C_{(2,1)}+C_{(3,1)};\\
X_1C_{(2,1)}=C_{(3,1)}+C_{(1,4)}+2C_{(0,4)}+C_{(1,1)}+C_{(2,2)}+C_{(1,2)};\\
X_1C_{(2,0)}=C_{(3,0)}+C_{(1,3)}+2C_{(0,3)}+X_1;\\
X_1C_{(1,3)}=C_{(2,3)}+2C_{(0,6)}+2C_{(3,0)}+2C_{(2,0)}+C_{(1,3)}+2C_{(0,3)};\\
X_1C_{(1,2)}=C_{(2,2)}+2C_{(0,5)}+C_{(1,2)}+2C_{(0,2)}+C_{(2,1)}+C_{(1,1)};\\
X_1C_{(1,1)}=C_{(2,1)}+2C_{(0,4)}+C_{(1,2)}+2C_{(0,2)}+C_{(1,1)}+2X_2;\\
X_1C_{(0,3)}=C_{(1,3)}+2C_{(2,0)}+2X_1;\\
X_1C_{(0,2)}=C_{(1,2)}+C_{(1,1)}+2X_2;\\
X_1X_1=C_{(2,0)}+2C_{(0,3)}+2X_1+6;\\
X_1X_2=C_{(1,1)}+2C_{(0,2)}+2X_2.
\end{gather*}

%\vfill\eject
Specializing the second of the generic relations to either $a\in\{0,1\}$ or $b\in\{0,1,2\}$, we have
\begin{gather*}
X_2C_{(1,b)}=C_{(1,b+1)}+C_{(2,b-1)}+2C_{(0,b+2)}+C_{(2,b-2)}+2C_{(0,b+1)}+C_{(1,b-1)};\\
X_2C_{(0,b)}=C_{(0,b+1)}+C_{(1,b-1)}+C_{(1,b-2)}+C_{(0,b-1)};\\
%%%%%%%%%
%X_2C_{(a,2)}=C_{(a,3)}+C_{(a+1,1)}+C_{(a+1,1)}+C_{(a-1,4)}+2C_{(a+1,0)}+C_{(a-1,2)};\\
X_2C_{(a,1)}=C_{(a,2)}+2C_{(a+1,0)}+C_{(a-1,3)}+C_{(a-1,2)}+2C_{(a,0)}+C_{(a,1)};\\
X_2C_{(a,0)}=C_{(a,1)}+C_{(a-1,2)}+C_{(a-1,1)};\\
%%%%%%%%%%
X_2C_{(1,2)}=C_{(1,3)}+C_{(2,1)}+2C_{(0,4)}+2C_{(0,3)}+C_{(1,1)}+2C_{(2,0)};\\
X_2C_{(1,1)}=C_{(1,2)}+2C_{(2,0)}+2C_{(0,3)}+2C_{(0,2)}+C_{(1,1)}+2X_1;\\
X_2C_{(0,2)}=C_{(0,3)}+C_{(1,1)}+2X_1+X_2;\\
X_2X_2=C_{(0,2)}+2X_1+2X_2+6.
\end{gather*}

%%%%%%%%%%%%%%%%%%%%%%%%%%%%%%%%%%%%%%%
\begin{table}[h]
\centering
\begin{tabular}{|l|l@{\,}|}
\hline
\multicolumn{2}{|c|}{$C$-polynomials}\\
\hline
$C_{(1,0)}$ & $X_1$\tsep{1pt}\bsep{1pt}\\
\hline
$C_{(0,1)}$ & $X_2$\tsep{1pt}\bsep{1pt}\\
\hline
$C_{(0,2)}$  &  $-6-2X_1-2X_2+X_2^2$\tsep{1pt}\bsep{1pt}\\
\hline
$C_{(1,1)}$  &  $12+4X_1+2X_2+X_1X_2-2X_2^2$\tsep{1pt}\bsep{1pt}\\
\hline
$C_{(1,2)}$  &  $-12-10X_1-2X_1^2-4X_2-3X_1X_2+2X_2^2+X_1X_2^2$\tsep{1pt}\bsep{1pt}\\
\hline
$C_{(0,3)}$  &  $-12-12X_1-2X_1^2-3X_2-3X_1X_2+2X_2^2+X_1X_2^2$\tsep{1pt}\bsep{1pt}\\
\hline
$C_{(2,0)}$  &  $18+22X_1+5X_1^2+6X_2+6X_1X_2-4X_2^2-2X_1X_2^2$\tsep{1pt}\bsep{1pt}\\
\hline
$C_{(1,3)}$  &  $ - 36 - 58X_1 - 22X_1^2 - 2X_1^3 - 12X_2 - 15X_1X_2 - 3X_1^2X_2 + 8X_2^2 + 6X_1X_2^2 + X_1^2X_2^2$\tsep{1pt}\bsep{1pt}\\
\hline
$C_{(0,4)}$  &  $6 + 8X_1 + 2X_1^2 - 8X_2 - 10X_1X_2 - 2X_1^2X_2 - 4X_2^2 - 4X_1X_2^2 + 2X_2^3 + X_1X_2^3$\tsep{1pt}\bsep{1pt}\\
\hline
$C_{(2,1)}$  &  $6X_1 + 2X_1^2 + 20X_2 + 24X_1X_2 + 5X_1^2X_2 + 6X_2^2 + 5X_1X_2^2 - 4X_2^3 - 2X_1X_2^3$\tsep{1pt}\bsep{1pt}\\
\hline
$C_{(1,4)}$  &  $ - 12 - 4X_1 + 6X_1^2 + 2X_1^3 - 22X_2 - 33X_1X_2 - 15X_1^2X_2 - 2X_1^3X_2 - 4X_2^2 - 9X_1X_2^2$\tsep{1pt}\bsep{1pt}\\
             &  $ - 4X_1^2X_2^2 + 4X_2^3 + 4X_1X_2^3 + X_1^2X_2^3$\bsep{1pt}\\
\hline
$C_{(3,0)}$  &  $60 + 99X_1 + 48X_1^2 + 7X_1^3 + 18X_2 + 27X_1X_2 + 9X_1^2X_2 - 12X_2^2 - 12X_1X_2^2 - 3X_1^2X_2^2$\tsep{1pt}\bsep{1pt}\\
\hline
$C_{(2,2)}$  &  $ - 108 - 156X_1 - 46X_1^2 - 2X_1^3 - 64X_2 - 60X_1X_2 + 9X_1^2X_2 + 5X_1^3X_2 + 32X_2^2 + 36X_1X_2^2$\tsep{1pt}\bsep{1pt}\\
             &  $ + 11X_1^2X_2^2 + 12X_2^3 + 6X_1X_2^3 - 2X_1^2X_2^3 - 4X_2^4 - 2X_1X_2^4$\bsep{1pt}\\
\hline
$C_{(3,1)}$  &  $108 + 150X_1 + 44X_1^2 + 2X_1^3 + 104X_2 + 135X_1X_2 + 34X_1^2X_2 + 2X_1^3X_2 - 20X_2^2$\tsep{1pt}\bsep{1pt}\\
             &  $ - 14X_1X_2^2 - 2X_1^2X_2^2 - 20X_2^3 - 16X_1X_2^3 - X_1^2X_2^3 + 4X_2^4 + 2X_1X_2^4$\bsep{1pt}\\
\hline
$C_{(4,0)}$  &  $198 + 400X_1 + 282X_1^2 + 84X_1^3 + 9X_1^4 + 240X_2 + 360X_1X_2 + 152X_1^2X_2 + 20X_1^3X_2$\tsep{1pt}\bsep{1pt}\\
             &  $ + 26X_2^2 + 28X_1X_2^2 - 8X_1^2X_2^2 - 4X_1^3X_2^2 - 52X_2^3 - 46X_1X_2^3 - 8X_1^2X_2^3$\bsep{1pt}\\
             &  $ - 8X_2^4 - 8X_1X_2^4 + 4X_2^5 + 2X_1X_2^5$\bsep{1pt}\\
\hline
\end{tabular}
\caption{The irreducible $C_{(a,b)}$-polynomials of $G_2$ with $a+b\leqslant4$.} \label{tableG2-Cpoly}
\end{table}
%%%%%%%%%%%%%%%%%%%%%%%%%%%%%%%%%%%%%%%

\begin{remark}
It can be seen from Table~\ref{tableG2-Cpoly} that order of $C_{(a,b)}$-polynomial sometimes exceeds $a+b$.
% It can be shown that it is equal to the integer part $[\tfrac85a+b]$.
\end{remark}

\subsection{Recursion relations for $\boldsymbol{S}$-functions of $\boldsymbol{G_2}$}
%%%%%%%%%%%%%%%%%%%%%%%%%%%%%%%%%%%%%%%%%%%%%%%%%%%%%%
Generic recursion relations for $S$-polynomials differ very little from those for $C$-polynomials.
\begin{gather*}
\text{for }a{\geqslant}3,\ b{\geqslant}4:
\\
X_1S_{(a,b)} = S_{(a + 1,b)} + S_{(a - 1,b + 3)} + S_{(a + 2,b - 3)} + S_{(a - 2,b + 3)}
        + S_{(a + 1,b - 3)} + S_{(a - 1,b)};
\\[1ex]
\text{for }a{\geqslant}2,\ b{\geqslant}3:
\\
X_2S_{(a,b)} = S_{(a,b + 1)} + S_{(a + 1,b - 1)} + S_{(a - 1,b + 2)} + S_{(a + 1,b - 2)}
        + S_{(a - 1,b + 1)} + S_{(a,b - 1)}.
\end{gather*}

The $S$-polynomials need not be calculated independently. They can be read off the tables~\cite{BMP} as the characters of $G_2$ representations.
The dimension $d_{(a,b)}$ of an irreducible representation of $G_2$ with the highest weight $\lambda=(a,b)$ is given by  \eqref{dimensionG2}.

Here are all $G_2$-characters $\chi_{(a,b)}$ with $a+b\leqslant3$:
\begin{gather*}
\chi_{(1,0)} = 1+C_{(1,0)} = 1+X_1;\\
\chi_{(0,1)} = 2+C_{(1,0)}+C_{(0,1)} = 2+X_1+X_2;\\
\chi_{(2, 0)}  =  3 + 2 X_1 + X_2 + C_{(2, 0)};\\
\chi_{(1, 1)}  =  4 + 4 X_1 + 2 X_2 + 2 C_{(2, 0)} + C_{(1, 1)};\\
\chi_{(3, 0)}  =  5 + 4 X_1 + 3 X_2 + 2 C_{(2, 0)} + C_{(3, 0)};\\
\chi_{(0, 2)}  =  5 + 3 X_1 + 3 X_2 + 2 C_{(2, 0)} + C_{(1, 1)} + C_{(3, 0)} + C_{(0, 2)};\\
\chi_{(2, 1)}  =  9 + 8 X_1 + 6 X_2 + 5 C_{(2, 0)} + 3 C_{(1, 1)} + 2 C_{(3, 0)} + C_{(0, 2)} + C_{(2, 1)};\\
\chi_{(1, 2)}  =  10 + 10X_1 + 7X_2 + 7C_{(2,0)} + 5C_{(1,1)}+ 3C_{(3,0)} + 3C_{(3,0)} + 2C_{(0,2)} + {}\\
\phantom{\chi_{(1, 2)}  =}
 2C_{(2,1)} + C_{(4,0)} + C_{(1,2)};\\
\chi_{(0, 3)}  =  9  +  7 X_1  +  7 X_2  +  5 C_{(2, 0)}  +  4 C_{(1, 1)}  +  4 C_{(3, 0)}  +  3 C_{(0, 2)}  +  2 C_{(1, 1)} +{} \\
\phantom{\chi_{(0, 3)}  = }C_{(4, 0)}  + C_{(1, 2)}  +  C_{(3, 1)}  +  C_{(0, 3)}.
\end{gather*}

%%%%%%%%%%%%%%%%%%%%%%%%%%%%%%%%%%%%%%%%%%%%%%%%%%%
\begin{table}
\centering
\begin{tabular}{|l|l|}
\hline
\multicolumn{2}{|c|}{$S$-polynomials}\\
\hline
$S_{(2,1)}$  &  $1 + X_1$$\phantom{\sum^{A^1}}$\\
\hline
$S_{(1,2)}$  &  $2 + X_1 + X_2$$\phantom{\sum^{A^1}}$\\
\hline
$S_{(3,1)}$  &  $21 + 24X_1 + 5X_1^2 + 7X_2 + 6X_1X_2 - 4X_2^2 - 2X_1X_2^2$$\phantom{\sum^{A^1}}$\\
\hline
$S_{(2,2)}$  &  $52 + 52X_1 + 10X_1^2 + 16X_2 + 13X_1X_2 - 10X_2^2 - 4X_1X_2^2$$\phantom{\sum^{A^1}}$\\
\hline
$S_{(4,1)}$  &  $113 + 151X_1 + 58X_1^2 + 7X_1^3 + 35X_2 + 40X_1X_2 + 9X_1^2X_2 - 22X_2^2 - 16X_1X_2^2 - 3X_1^2X_2^2$$\phantom{\sum^{A^1}}$\!\!\!\!\!\!\!\\
\hline
$S_{(1,3)}$  &  $107 + 148X_1 + 58X_1^2 + 7X_1^3 + 33X_2 + 40X_1X_2 + 9X_1^2X_2 - 21X_2^2 - 16X_1X_2^2 - 3X_1^2X_2^2$$\phantom{\sum^{A^1}}$\!\!\!\!\!\!\!\\
\hline
$S_{(3,2)}$  &  $249 + 332X_1 + 123X_1^2 + 14X_1^3 + 96X_2 + 111X_1X_2 + 23X_1^2X_2 - 43X_2^2 - 29X_1X_2^2$$\phantom{\sum^{A^1}}$\\
             &  $ - 6X_1^2X_2^2 - 4X_2^3 - 2X_1X_2^3$\\
\hline
$S_{(2,3)}$  &  $550 + 879X_1 + 463X_1^2 + 105X_1^3 + 9X_1^4 + 385X_2 + 533X_1X_2 + 189X_1^2X_2 + 20X_1^3X_2$$\phantom{\sum^{A^1}}$\\
             &  $ - 11X_1X_2^2 - 32X_2^2 - 17X_1^2X_2^2  - 4X_1^3X_2^2 - 60X_2^3 - 50X_1X_2^3 - 8X_1^2X_2^3 - 8X_2^4$\\
             &  $ - 8X_1X_2^4 + 4X_2^5 + 2X_1X_2^5$\\
\hline
$S_{(4,1)}$  &  $651 + 1063X_1 + 543X_1^2 + 114X_1^3 + 9X_1^4 + 488X_2 + 679X_1X_2 + 232X_1^2X_2$$\phantom{\sum^{A^1}}$\\
             &  $ + 22X_1^3X_2 - 32X_1X_2^2 - 51X_2^2 - 22X_1^2X_2^2 - 4X_1^3X_2^2 - 80X_2^3 - 66X_1X_2^3 - 9X_1^2X_2^3$\\
             &  $ - 4X_2^4 - 6X_1X_2^4 + 4X_2^5 + 2X_1X_2^5$\\
\hline
\end{tabular}
\caption{The irreducible $S_{(a,b)}$-polynomials of $G_2$ with $a+b\leqslant5$.} \label{tableG2-Spoly}
\end{table}
%%%%%%%%%%%%%%%%%%%%%%%%%%%%%%%%%%%%%%%%%%%%%%%%%%%

%%%%%%%%%%%%%%%%%%%%%%%%%%%%%%%%%%%%%%%%%%%%%%%%%%%%%%%%
\section{Recursion relations for Lie algebras of rank 3}\label{sec_3D}

\subsection{Recursion relations for $\boldsymbol{C}$-functions of $\boldsymbol{A_3}$}

There are 4 congruence classes of $A_3$ defined by
\begin{gather}
\#(a,b,c)=a+2b+3c\mod4.
\end{gather}
The dimension $d_{(a,b,c)}$ of the irreducible representation of $A_3$ with the highest weight \mbox{$\lambda=(a,b,c)$} is given by
\begin{gather}
d_{(a,b,c)}=\tfrac1{12}(a+1)(b+1)(c+1)(a+b+2)(b+c+2)(a+b+c+3).
\end{gather}

The variables of the $A_3$ polynomials are chosen to be
\begin{gather*}\label{3variblesbles}
X_1:=C_{(1,0,0)}(x_1,x_2,x_3),\qquad
X_2:=C_{(0,1,0)}(x_1,x_2,x_3),\qquad
X_3:=C_{(0,0,1)}(x_1,x_2,x_3).
\end{gather*}
The orbit functions $C_{(1,0,0)}$ and $C_{(0,0,1)}$ contain 4 exponential terms, and $C_{(0,1,0)}$ has 6 terms.
All~terms have the form $e^{2\pi i\l \mu,x\r}$, where $\mu$ runs over the po\-ints/weights of the corresponding orbit.

As for $C$-functions, generic recursion relations are the decompositions of the following products, where we assume~$a,b,c\geqslant2$:
\begin{gather*}
X_1C_{(a,b,c)}  = C_{(a + 1,b,c)} + C_{(a - 1,b + 1,c)} + C_{(a,b - 1,c + 1)} + C_{(a,b,c - 1)};\\
X_2C_{(a,b,c)}  = C_{(a,b + 1,c)} + C_{(a + 1,b - 1,c + 1)} + C_{(a - 1,b,c + 1)} + C_{(a + 1,b,c - 1)} +{} \\
\phantom{X_2C_{(a,b,c)}  = }C_{(a - 1,b + 1,c - 1)} + C_{(a,b - 1,c)};\\
X_3C_{(a,b,c)}  = C_{(a,b,c + 1)} + C_{(a,b + 1,c - 1)} + C_{(a + 1,b - 1,c)} + C_{(a - 1,b,c)}.
\end{gather*}
Note that the first and the third relations are easily obtained from each other by interchanging the first and third component of all dominant weights.
Thus
\begin{gather*}
X_1\leftrightarrow X_3\qquad\text{and}\qquad(a,b,c) \leftrightarrow\ (c,b,a).
\end{gather*}

The special recursion relations are obtained from the same products, where some of the components $a$, $b$, $c$ of the generic dominant weight take special values $1$ and $0$. In these cases, the decompositions of the three products are modified. To have a complete set of recursion relations, every combination of $a$, $b$, $c$ with values 0 and 1 needs to be used.

Solving the following additional recursion relations, we obtain some lowest $C$-polynomials of~$A_3$ given in Table~\ref{tableA3-CSpoly}.
\begin{gather*}
X_1C_{(a,b,1)}=2C_{(a,b,0)}+C_{(a+1,b,1)}+C_{(a-1,b+1,1)}+C_{(a,b-1,2)};\\
X_1C_{(a,b,0)}=C_{(a+1,b,0)}+C_{(a-1,b+1,0)}+C_{(a,b-1,1)};\\
X_1C_{(a,1,c)}=C_{(a+1,1,c)}+C_{(a-1,2,c)}+C_{(a,1,c-1)}+2C_{(a,0,c+1)};\\
X_1C_{(a,1,1)}=2C_{(a,1,0)}+C_{(a+1,1,1)}+C_{(a-1,2,1)}+2C_{(a,0,2)};\\
X_1C_{(a,1,0)}=C_{(a+1,1,0)}+C_{(a-1,2,0)}+2C_{(a,0,1)};\\
X_1C_{(a,0,c)}=C_{(a+1,0,c)}+C_{(a-1,1,c)}+C_{(a,0,c-1)};\\
X_1C_{(a,0,1)}=3C_{(a,0,0)}+C_{(a+1,0,1)}+C_{(a-1,1,1)};\\
X_1C_{(a,0,0)}=C_{(a+1,0,0)}+C_{(a-1,1,0)};\\
X_1C_{(1,b,c)}=C_{(2,b,c)}+2C_{(0,b+1,c)}+C_{(1,b,c-1)}+C_{(1,b-1,c+1)};\\
X_1C_{(1,b,1)}=2C_{(1,b,0)}+C_{(2,b,1)}+2C_{(0,b+1,1)}+C_{(1,b-1,2)};\\
X_1C_{(1,b,0)}=2C_{(0,b+1,0)}+C_{(2,b,0)}+C_{(1,b-1,1)};\\
X_1C_{(1,1,c)}=C_{(2,1,c)}+2C_{(0,2,c)}+C_{(1,1,c-1)}+2C_{(1,0,c+1)};\\
X_1C_{(1,1,1)}=2C_{(1,1,0)}+C_{(2,1,1)}+2C_{(0,2,1)}+2C_{(1,0,2)};\\
X_1C_{(1,1,0)}=2C_{(0,2,0)}+C_{(2,1,0)}+2C_{(1,0,1)};\\
X_1C_{(1,0,c)}=C_{(2,0,c)}+2C_{(0,1,c)}+C_{(1,0,c-1)};\\
X_1C_{(1,0,1)}=3X_1+2C_{(0,1,1)}+C_{(2,0,1)};\\
X_1C_{(0,b,c)}=C_{(1,b,c)}+C_{(0,b,c-1)}+C_{(0,b-1,c+1)};\\
X_1C_{(0,b,1)}=2C_{(0,b,0)}+C_{(1,b,1)}+C_{(0,b-1,2)};\\
X_1C_{(0,b,0)}=C_{(1,b,0)}+C_{(0,b-1,1)};\\
X_1C_{(0,1,c)}=2C_{(0,0,c+1)}+C_{(1,1,c)}+C_{(0,1,c-1)}+C_{(0,0,c+1)};\\
X_1C_{(0,1,1)}=2X_2+3C_{(0,0,2)}+C_{(1,1,1)};\\
X_1C_{(0,0,c)}=C_{(1,0,c)}+C_{(0,0,c-1)};\\
X_1^2=C_{(2,0,0)}+2X_2;\quad
X_1X_2=C_{(1,1,0)}+3X_3;\quad
X_1X_3=4+C_{(1,0,1)};\\
%%%%
X_2C_{(a,b,1)}=2C_{(a+1,b,0)}+2C_{(a-1,b+1,0)}+C_{(a,b-1,1)}+C_{(a,b+1,1)}+   C_{(a-1,b,2)}+C_{(a+1,b-1,2)};
\\
X_2C_{(a,b,0)}=C_{(a,b-1,0)}+C_{(a,b+1,0)}+C_{(a-1,b,1)}+C_{(a+1,b-1,1)};\\
X_2C_{(a,1,c)}=2C_{(a,0,c)}+2C_{(a+1,0,c+1)}+C_{(a-1,1,c+1)}+C_{(a-1,2,c-1)}+
            C_{(a+1,1,c-1)}+C_{(a,2,c)};\\
X_2C_{(a,1,1)}=2C_{(a+1,1,0)}+2C_{(a-1,2,0)}+2C_{(a,0,1)}+C_{(a,2,1)}+
      2C_{(a+1,0,2)}+C_{(a-1,1,2)};\\
X_2C_{(a,1,0)}=3C_{(a,0,0)}+C_{(a,2,0)}+2C_{(a+1,0,1)}+C_{(a-1,1,1)};\\
X_2C_{(a,0,c)}=C_{(a,1,c)}+C_{(a+1,0,c-1)}+C_{(a-1,2,c-1)}+C_{(a-1,0,c+1)};\\
X_2C_{(a,0,1)}=3C_{(a+1,0,0)}+2C_{(a-1,1,0)}+C_{(a,1,1)}+C_{(a-1,0,2)};\\
X_2C_{(a,0,0)}=C_{(a,1,0)}+C_{(a-1,0,1)};\\
X_2C_{(1,b,c)}=C_{(1,b-1,c)}+C_{(1,b+1,c)}+C_{(2,b,c-1)}+2C_{(0,b+1,c-1)}+
            2C_{(0,b,c+1)}+C_{(2,b-1,c+1)};\\
X_2C_{(1,b,1)}=4C_{(0,b+1,0)}+2C_{(2,b,0)}+C_{(1,b-1,1)}+C_{(1,b+1,1)}+
            2C_{(0,b,2)}+C_{(2,b-1,2)};\\
X_2C_{(1,b,0)}=C_{(1,b-1,0)}+C_{(1,b+1,0)}+2C_{(0,b,1)}+C_{(2,b-1,1)};\\
X_2C_{(1,1,c)}=2C_{(1,0,c)}+C_{(1,2,c)}+C_{(2,1,c-1)}+2C_{(0,2,c-1)}+
         2C_{(2,0,c+1)}+2C_{(0,1,c+1)};\\
X_2C_{(1,1,1)}=4C_{(0,2,0)}+2C_{(2,1,0)}+2C_{(1,0,1)}+C_{(1,2,1)}+2C_{(2,0,2)}+2C_{(0,1,2)};\\
X_2C_{(1,1,0)}=3X_1+C_{(1,2,0)}+2C_{(2,0,1)}+2C_{(0,1,1)};\\
X_2C_{(1,0,c)}=3C_{(0,0,c+1)}+C_{(1,1,c)}+C_{(2,0,c-1)}+2C_{(0,1,c-1)};\\
X_2C_{(1,0,1)}=3C_{(2,0,0)}+4C_{(0,2,0)}+3C_{(0,0,2)}+C_{(1,1,1)};\\
X_2C_{(0,b,c)}=C_{(0,b-1,c)}+C_{(0,b+1,c)}+C_{(1,b,c-1)}+C_{(1,b-1,c+1)};\\
X_2C_{(0,b,1)}=2C_{(1,b,0)}+C_{(0,b-1,1)}+C_{(0,b+1,1)}+C_{(1,b-1,2)};\\
X_2C_{(0,b,0)}=C_{(0,b+1,0)}+C_{(1,b-1,1)}+C_{(0,b-1,0)};\\
X_2C_{(0,1,c)}=C_{(0,2,c)}+C_{(1,1,c-1)}+2C_{(1,0,c+1)}+3C_{(0,0,c)};\\
X_2C_{(0,1,1)}=2C_{(1,1,0)}+3X_3+C_{(0,2,1)}+2C_{(1,2,0)};\\
X_2C_{(0,0,c)}=C_{(0,1,c)}+C_{(1,0,c-1)};\\
X_2^2=6+C_{(0,2,0)}+2C_{(1,0,1)};\quad
X_2X_3=3X_1+C_{(0,1,1)};\\
%%%
X_3C_{(a,b,1)}=2C_{(a,b+1,0)}+C_{(a-1,b,1)}+C_{(a+1,b-1,1)}+C_{(a,b,2)};\\
X_3C_{(a,b,0)}=C_{(a-1,b,0)}+C_{(a+1,b-1,0)}+C_{(a,b,1)};\\
X_3C_{(a,1,c)}=2C_{(a+1,0,c)}+C_{(a-1,1,c)}+C_{(a,2,c-1)}+C_{(a,1,c+1)};\\
X_3C_{(a,1,1)}=2C_{(a,2,0)}+2C_{(a+1,0,1)}+C_{(a-1,1,1)}+C_{(a,1,2)};\\
X_3C_{(a,1,0)}=3C_{(a+1,0,0)}+C_{(a-1,1,0)}+C_{(a,1,1)};\\
X_3C_{(a,0,c)}=C_{(a-1,0,c)}+C_{(a,1,c-1)}+C_{(a,0,c+1)};\\
X_3C_{(a,0,1)}=2C_{(a,1,0)}+C_{(a-1,0,1)}+C_{(a,0,2)};\\
X_3C_{(a,0,0)}=C_{(a-1,0,0)}+C_{(a,0,1)};\\
X_3C_{(1,b,c)}=C_{(1,b,c+1)}+C_{(1,b+1,c-1)}+2C_{(0,b,c)}+C_{(2,b-1,c)};\\
X_3C_{(1,b,1)}=2C_{(1,b+1,0)}+2C_{(0,b,1)}+C_{(2,b-1,1)}+C_{(1,b,2)};\\
X_3C_{(1,b,0)}=2C_{(0,b,0)}+C_{(2,b-1,0)}+C_{(1,b,1)};\\
X_3C_{(1,1,c)}=2C_{(0,1,c)}+C_{(1,2,c-1)}+C_{(1,1,c+1)}+2C_{(2,0,c)};\\
X_3C_{(1,1,1)}=2C_{(1,2,0)}+2C_{(2,0,1)}+2C_{(0,1,1)}+C_{(1,1,2)};\\
X_3C_{(1,1,0)}=3C_{(2,0,0)}+2X_2+C_{(1,1,1)};\\
X_3C_{(1,0,c)}=2C_{(0,0,c)}+C_{(1,1,c-1)};\\
X_3C_{(1,0,1)}=2C_{(1,1,0)}+3X_3+C_{(1,0,2)};\\
X_3C_{(0,b,c)}=C_{(1,b-1,c)}+C_{(0,b+1,c-1)}+C_{(0,b,c+1)};\\
X_3C_{(0,b,1)}=2C_{(0,b+1,0)}+C_{(1,b-1,1)}+C_{(0,1,2)};\\
X_3C_{(0,b,0)}=C_{(1,b-1,0)}+C_{(0,b,1)};\\
X_3C_{(0,1,c)}=C_{(1,0,c)}+C_{(0,2,c-1)}+C_{(0,1,c+1)}+C_{(1,0,c)};\\
X_3C_{(0,1,1)}=C_{(0,1,2)}+2C_{(0,2,0)}+2C_{(1,0,1)};\\
X_3C_{(0,0,c)}=C_{(0,0,c+1)}+C_{(0,1,c-1)};\qquad
X_3^2=C_{(0,0,2)}+2X_2.
\end{gather*}

\subsection{$\boldsymbol{S}$-polynomials of $\boldsymbol{A_3}$}
%%%%%%%%%%%%%%%%%%%%%%%%%%%%%%%%%%%%%%%%%%%%%%%%%%%%%%
We use the notation $S=S_{(1,1,1)}(x_1,x_2,x_3)$. It is the sum of 24 exponential terms. Generic recursion relations are decompositions of the following products, where we assume that \mbox{$a,b,c>1$}:
\begin{gather*}
X_1S_{(a,b,c)}  = S_{(a + 1,b,c)} + S_{(a - 1,b + 1,c)} + S_{(a,b - 1,c + 1)} + S_{(a,b,c - 1)};\\
X_2S_{(a,b,c)}  = S_{(a,b + 1,c)} + S_{(a + 1,b - 1,c + 1)} + S_{(a - 1,b,c + 1)} + S_{(a + 1,b,c - 1)} +   S_{(a - 1,b + 1,c - 1)} + S_{(a,b - 1,c)};\\
X_3S_{(a,b,c)}  = S_{(a,b,c + 1)} + S_{(a,b + 1,c - 1)} + S_{(a + 1,b - 1,c)} + S_{(a - 1,b,c)}.
\end{gather*}

The terms in the decomposition of special recursion relations for $S$-functions differ from those of $C$-functions if some $S_{(a,b,c)}(x_1,x_2,x_3)=0$. This occurs if some of its dominant weight components $a$, $b$, $c$ equal zero.

To calculate $S$-polynomials explicitly (see Table~\ref{tableA3-CSpoly}) we use the $A_3$ characters.
The lowest ones from the congruence classes $\#=0$, $\#=1$, $\#=2$ and $\#=3$ are listed below:
\begin{gather*}
\begin{array}{@{}ll@{}}
\#=0\colon                                                     &\#{=}1\colon\\
\chi_{(0,0,0)} = C_{(0,0,0)} = 1,                              &\chi_{(1,0,0)} = C_{(1,0,0)} = X_{1},\\
\chi_{(1,0,1)} = 3 + C_{(1,0,1)},                              &\chi_{(0,1,1)} = 2X_{1} + C_{(0,1,1)},\\
\chi_{(0,2,0)} = 2 + C_{(1,0,1)} + C_{(0,2,0)},                &\chi_{(2,0,1)} = 3X_{1} + C_{(0,1,1)} + C_{(2,0,1)},\\
\chi_{(0,1,2)} = 3 + 2C_{(1,0,1)} + C_{(0,2,0)} + C_{(0,1,2)}, &\chi_{(0,0,3)} = X_{1} + C_{(0,1,1)} + C_{(0,0,3)},\\
\chi_{(2,1,0)} = 3 + 2C_{(1,0,1)} + C_{(0,2,0)} + C_{(2,1,0)};\  &\chi_{(1,2,0)} = 3X_1 + 2C_{(0,1,1)} + C_{(2,0,1)} + C_{(1,2,0)};
\\[1ex]
\#=2\colon                                                               &\#{=}3\colon\\
\chi_{(0,1,0)} = C_{(0,1,0)} = X_{2},                                    &\chi_{(0,0,1)} = C_{(0,0,1)} = X_{3},\\
\chi_{(0,0,2)} = X_{2} + C_{(0,0,2)},                                    &\chi_{(1,1,0)} = 2X_{3} + C_{(1,1,0)},\\
\chi_{(2,0,0)} = X_{2} + C_{(2,0,0)},                                    &\chi_{(1,0,2)} = 3X_{3} + C_{(1,1,0)} + C_{(1,0,2)},\\
\chi_{(1,1,1)} = 4X_{2} + 2C_{(0,0,2)} + 2C_{(2,0,0)} + C_{(1,1,1)};\ &\chi_{(3,0,0)} = X_{3} + C_{(1,1,0)} + C_{(3,0,0)},\\
&\chi_{(0,2,1)} = 3X_{3} + 2C_{(1,1,0)} + C_{(1,0,2)} + C_{(0,2,1)}.
\end{array}
\end{gather*}

  It should be mentioned that some additional information about polynomials connected with~$A_n$ can be found in~\cite{NPT}.
We believe that orthogonal polynomials of $A_n$ are natural $n$-dimensional generalizations of Chebyshev polynomials,
but other approaches exist~\cite{Dunkl}.

\begin{table}[h]
\centering
\begin{tabular}{|l|l@{}|c|l|l@{}|}
\cline{1-2} \cline{4-5}
\multicolumn{2}{|c|}{$C$-polynomials}&\quad\quad& \multicolumn{2}{|c|}{$S$-polynomials$\phantom{\sum^{A^1}}$}\\
\cline{1-2}\cline{4-5}
\multicolumn{2}{|l|}{$\#=0$}&\quad\quad& \multicolumn{2}{|l|}{$\#=0$$\phantom{\sum^{A^1}}$}\\
\cline{1-2}\cline{4-5}
$C_{(1,0,1)}$&$ - 4 + X_{1}X_{3}$                                   &&$S_{(2,1,2)}$&$ - 1 + X_1X_3$$\phantom{\sum^{A^1}}$\\
\cline{1-2}\cline{4-5}
$C_{(0,2,0)}$&$2 - 2X_{1}X_{3} + X_{2}^{2}$                         &&$S_{(1,3,1)}$&$X_2^2 - X_1X_3$$\phantom{\sum^{A^1}}$\\
\cline{1-2}\cline{4-5}
$C_{(0,1,2)}$&$4 - X_{1}X_{3} - 2X_{2}^{2} + X_{2}X_{3}^{2}$          &&$S_{(1,2,3)}$&$1 - X_2^2 - X_1X_3 + X_2X_3^2$$\phantom{\sum^{A^1}}$\\
\cline{1-2}\cline{4-5}
$C_{(2,1,0)}$&$4 - X_{1}X_{3} + X_{1}^{2}X_{2} - 2X_2^2$              &&$S_{(3,2,1)}$&$1 + X_1^2X_2 - X_2^2 - X_1X_3$$\phantom{\sum^{A^1}}$\\
\cline{1-2}\cline{4-5}
\multicolumn{2}{|l|}{$\#=1$}&\quad\quad& \multicolumn{2}{|l|}{$\#=1$$\phantom{\sum^{A^1}}$}\\
\cline{1-2}\cline{4-5}
$C_{(1,0,0)}$&$X_1$                                             &&$S_{(2,1,1)}$&$X_1$$\phantom{\sum^{A^1}}$\\
\cline{1-2}\cline{4-5}
$C_{(0,1,1)}$&$-3X_{1}+X_{2}X_{3}$                              &&$S_{(1,2,2)}$&$-X_1+X_2X_3$$\phantom{\sum^{A^1}}$\\
\cline{1-2}\cline{4-5}
$C_{(0,0,3)}$&$3X_{1}-3X_{2}X_{3}+X_{3}^{3}$                    &&$S_{(1,1,4)}$&$X_1-2X_2X_3+X_3^3$$\phantom{\sum^{A^1}}$\\
\cline{1-2}\cline{4-5}
$C_{(2,0,1)}$&$-X_{1}-2X_{2}X_{3}+X_{1}^{2}X_{3}$               &&$S_{(3,1,2)}$&$-X_1+X_1^2X_3-X_2X_3$$\phantom{\sum^{A^1}}$\\
\cline{1-2}\cline{4-5}
$C_{(1,2,0)}$&$5X_{1} - X_{2}X_{3} - 2X_{1}^{2}X_{3} + X_{1}X_{2}^{2}$$\phantom{\sum^{A^1}}$&&$S_{(2,3,1)}$&$X_1 + X_1X_2^2 - X_1^2X_3 - X_2X_3$$\phantom{\sum^{A^1}}$\\
\cline{1-2}\cline{4-5}
\multicolumn{2}{|l|}{$\#=2$}&\quad\quad& \multicolumn{2}{|l|}{$\#=2$$\phantom{\sum^{A^1}}$}\\
\cline{1-2}\cline{4-5}
$C_{(0,1,0)}$&$X_2$                                             &&$S_{(1,2,1)}$&$X_2$$\phantom{\sum^{A^1}}$\\
\cline{1-2}\cline{4-5}
$C_{(0,0,2)}$&$-2X_{2}+X_{3}^{2}$                               &&$S_{(1,1,3)}$&$-X_2+X_3^2$$\phantom{\sum^{A^1}}$\\
\cline{1-2}\cline{4-5}
$C_{(2,0,0)}$&$-2X_{2}+X_{1}^{2}$                               &&$S_{(3,1,1)}$&$X_1^2-X_2$$\phantom{\sum^{A^1}}$\\
\cline{1-2}\cline{4-5}
$C_{(1,1,1)}$&$4X_{2} - 3X_{3}^{2} - 3X_{1}^{2} + X_1X_2X_3$          &&$S_{(2,2,2)}$&$-X_1^2-X_3^2+X_1X_2X_3$$\phantom{\sum^{A^1}}$\\
\cline{1-2}\cline{4-5}
\multicolumn{2}{|l|}{$\#=3$}&\quad\quad& \multicolumn{2}{|l|}{$\#=3$$\phantom{\sum^{A^1}}$}\\
\cline{1-2}\cline{4-5}
$C_{(0,0,1)}$&$X_3$                                             &&$S_{(1,1,2)}$&$X_3$$\phantom{\sum^{A^1}}$\\
\cline{1-2}\cline{4-5}
$C_{(1,1,0)}$&$-3X_{3}+X_{1}X_{2}$                              &&$S_{(2,2,1)}$&$X_1X_2-X_3$$\phantom{\sum^{A^1}}$\\
\cline{1-2}\cline{4-5}
$C_{(3,0,0)}$&$3X_{3}-3X_{1}X_{2}+X_{1}^{3}$                    &&$S_{(4,1,1)}$&$X_1^3-X_1X_2+X_3$$\phantom{\sum^{A^1}}$\\
\cline{1-2}\cline{4-5}
$C_{(1,0,2)}$&$-X_{3}-2X_{1}X_{2}+X_{1}X_{3}^{2}$               &&$S_{(2,1,3)}$&$-X_1X_2-X_3+X_1X_3^2$$\phantom{\sum^{A^1}}$\\
\cline{1-2}\cline{4-5}
$C_{(0,2,1)}$&$5X_{3} - X_{1}X_{2} - 2X_{1}X_{3}^{2} + X_{2}^{2}X_{3}$ &&$S_{(1,3,2)}$&$ - X_1X_2 - X_3 - X_1X_3^2 + X_2^2X_3$$\phantom{\sum^{A^1}}$\hspace{-100pt}\\
\cline{1-2} \cline{4-5}
\end{tabular}
\caption{The irreducible $C$-polynomials and $S$-polynomials of $A_3$.}\label{tableA3-CSpoly}
\end{table}

\subsection{Recursion relations for $\boldsymbol{C}$- and $\boldsymbol{S}$-polynomials of $\boldsymbol{B_3}$ and  $\boldsymbol{C_3}$}
%%%%%%%%%%%%%%%%%%%%%%%%%%%%%%%%%%%%%%%%%%%%%%%%%%%%%%%%
The two cases differ in many important respects in spite of the isomorphism of their Weyl groups.

We write the generic relations for the $C$-polynomials of the Lie algebras $B_3$ and $C_3$, respectively of the simple Lie group $O(7)$ and $Sp(6)$. The generic relations for the $S$-polynomials are obtained by replacing the $C$ symbol by $S$.

The variables are denoted by the same symbols $X_1$, $X_2$, $X_3$ for all algebras of rank 3, namely $X_j:=C_{\omega_j}$, $j=1,2,3$.
As orbits, they differ for different algebras. They are real-valued for $B_3$ and $C_3$, and complex for $A_3$.

There are two congruence classes of $(a_1,a_2,a_3)$ for either of the two algebras.

We have
\begin{gather*}
\#(B_3)=a_3\mod2\qquad\text{and}\qquad\#(C_3)=a_1+a_3\mod2.
\end{gather*}
The congruence numbers add up in a product.
For example, in the case of $B_3$
\begin{gather*}
{\#}(X_1S_{(a,b,c)})  = {\#}(S_{(a,b,c)}),\qquad
{\#}(X_2S_{(a,b,c)})  = {\#}(S_{(a,b,c)}),\\
{\#}(X_3S_{(a,b,c)})  = {\#}(S_{(a,b,c)}) + 1,
\end{gather*}
while for $C_3$, we have
\begin{gather*}
{\#}(X_1S_{(a,b,c)})  = {\#}(S_{(a,b,c)}) + 1,\qquad
{\#}(X_2S_{(a,b,c)})  = {\#}(S_{(a,b,c)}),\\
{\#}(X_3S_{(a,b,c)})  = {\#}(S_{(a,b,c)}) + 1.
\end{gather*}

The dimensions of irreducible representations are given by
\begin{gather*}
d_{(a,b,c)}(B_3)=\tfrac1{720}(a + 1)(b + 1)(c + 1)(a + b + 2)(2b + c + 3)(2a + 2b + 5)\times{}\\
\phantom{d_{(a,b,c)}(B_3)=}(b + c + 2)(a + b + c + 3)(a + 2b + c + 4);\\
d_{(a,b,c)}(C_3)= \tfrac1{720}(a + 1)(b + 1)(c + 1)(a + b + 2)(b + 2c + 3)(a + b + 2c + 4)\times{}\\
\phantom{d_{(a,b,c)}(C_3)=}(b + c + 2)(a + 2b + 2c + 5)(a + b + c + 3).
\end{gather*}

For $B_3$, we have
\begin{gather*}
X_1C_{(a,b,c)}=C_{(a + 1,b,c)} + C_{(a - 1,b + 1,c)} + C_{(a,b - 1,c + 2)} + C_{(a,b + 1,c - 2)} + C_{(a + 1,b - 1,c)} + C_{(a - 1,b,c)},\\
\phantom{X_1C_{(a,b,c)} = }\text{for}\quad a,b \geqslant 2,\ c\geqslant3;\\
X_2C_{(a,b,c)}=C_{(a,b + 1,c)} + C_{(a + 1,b - 1,c + 2)} + C_{(a - 1,b,c + 2)} + C_{(a + 1,b + 1,c - 2)} + C_{(a - 1,b + 2,c - 2)} + \\
\phantom{X_2C_{(a,b,c)} = }C_{(a + 2,b - 1,c)} + C_{(a + 1,b - 2,c + 2)} + C_{(a - 2,b + 1,c)} + C_{(a - 1,b - 1,c + 2)} + C_{(a + 1,b,c - 2)} + \\
\phantom{X_2C_{(a,b,c)} = }C_{(a - 1,b + 1,c - 2)} + C_{(a,b - 1,c)},\qquad
a \geqslant 2,\ b,c\geqslant3;\\
X_3C_{(a,b,c)}=C_{(a,b,c + 1)} + C_{(a,b + 1,c - 1)} + C_{(a + 1,b - 1,c + 1)} + C_{(a - 1,b,c + 1)} - C_{(a + 1,b,c - 1)} + \\
\phantom{X_3C_{(a,b,c)} = }C_{(a - 1,b + 1,c - 1)} + C_{(a,b - 1,c + 1)} + C_{(a,b,c - 1)},\quad
a,b,c\geqslant2.
\end{gather*}

For $C_3$, we have the generic recursion relations
\begin{gather*}
X_1C_{(a,b,c)}=C_{(a+1,b,c)}+C_{(a - 1,b+1,c)}+C_{(a,b - 1,c+1)}+C_{(a,b+1,c - 1)}+\\
\phantom{X_1C_{(a,b,c)}=}C_{(a+1,b - 1,c)}+C_{(a - 1,b,c)},
\qquad a,b,c\geqslant2;\\
X_2C_{(a,b,c)}=C_{(a,b+1,c)} + C_{(a+1,b - 1,c)}+C_{(a - 1,b,c+1)} + C_{(a+1,b+1,c - 1)} + C_{(a - 1,b+2,c - 1)} + \\
\phantom{X_2C_{(a,b,c)}=}C_{(a+2,b - 1,c)} + C_{(a+1,b - 2,c+1)} + C_{(a - 2,b+1,c)} + C_{(a+1,b,c - 1)} + C_{(a - 1,b - 1,c+1)} +  \\
\phantom{X_2C_{(a,b,c)}=} C_{(a - 1,b+1,c - 1)} + C_{(a,b - 1,c)},\qquad
a,b \geqslant 3,\ c\geqslant2;\\
X_3C_{(a,b,c)}=C_{(a,b,c+1)}+C_{(a,b+2,c - 1)}+C_{(a+2,b - 2,c+1)}+C_{(a - 2,b,c+1)}+C_{(a+2,b,c - 1)}+\\
\phantom{X_3C_{(a,b,c)}=}  C_{(a - 2,b+2,c - 1)}+C_{(a,b - 2,c+1)}+C_{(a,b,c - 1)},\qquad
a,b \geqslant 3,\ c\geqslant2.
\end{gather*}

\subsection{$C$- and $S$-polynomials of $B_3$ in three real variables}
For group $B_3$, our new coordinates $u$ satisfy the relation $u_i=\overline{u}_i$, $i=1,2,3$.

The following generic recursion relations for $C$-polynomials hold true when $k,l,m\geqslant 2$:
\begin{gather*}
C_{(k{+}1,l,m)}(u)=u_1C_{(k,l,m)}(u){-}C_{(k,l{+}1,m{-}2)}(u){-}C_{(k,l{-}1,m{+}2)}(u){-}C_{(k{+}1,l{-}1,m)}(u)\\
\phantom{C_{(k{+}1,l,m)}(u)=}                    {-}C_{(k{-}1,l{+}1,m)}(u){-}C_{(k{-}1,l,m)}(u){-}C_{(k{-}1,l{+}1,m)}(u){-}C_{(k{-}1,l,m)}(u),\\
C_{(k,l{+}1,m)}(u)=u_2C_{(k,l,m)}(u){-}C_{(k{+}1,l{-}1,m{+}2)}(u){-}C_{(k{-}1,l{-}1,m{+}2)}(u){-}C_{(k{+}1,l{-}2,m{+}2)}(u)\\
\phantom{C_{(k,l{+}1,m)}(u)=}                    {-}C_{(k{-}1,l,m{+}2)}(u){-}C_{(k{-}1,l{+}2,m{-}2)}(u){-}C_{(k{+}1,l{+}1,m{-}2)}(u)\\
\phantom{C_{(k,l{+}1,m)}(u)=}                    {-}C_{(k{-}1,l{+}1,m{-}2)}(u){-}C_{(k{+}1,l,m{-}2)}(u){-}C_{(k{-}2,l{+}1,m)}(u){-}C_{(k{+}2,l{-}1,m)}(u),\\
C_{(k,l,m{+}1)}(u)=u_3C_{(k,l,m)}(u){-}C_{(k{+}1,l{-}1,m{+}1)}(u){-}C_{(k,l{-}1,m{+}1)}(u){-}C_{(k{-}1,l,m{+}1)}(u)\\
\phantom{C_{(k,l,m{+}1)}(u)=}                    {-}C_{(k{-}1,l{+}1,m{-}1)}(u){-}C_{(k,l{+}1,m{-}1)}(u){-}C_{(k{+}1,l,m{-}1)}(u){-}C_{(k,l,m{-}1)}(u)
\end{gather*}

Remaining recurrence relations except for the lowest polynomials are listed below:
\begin{gather*}
C_{(k{+}1,l,0)}(u)=u_1C_{(k,l,0)}(u){-}C_{(k{-}1,l{+}1,0)}(u){-}C_{(k{+}1,l{-}1,0)}(u){-}C_{(k{-}1,l,0)}(u){-}C_{(k,l{-}1,2)}(u),\\
C_{(k,l{+}1,0)}(u)=u_2C_{(k,l,0)}(u){-}C_{(k{-}2,l{+}1,0)}(u){-}C_{(k{+}2,l{-}1,0)}(u){-}C_{(k{-}1,l,2)}(u){-}C_{(k,l{-}1,0)}(u)\\
\phantom{C_{(k,l{+}1,0)}(u)=}                    {-}C_{(k{+}1,l{-}1,2)}(u){-}C_{(k{-}1,l{-}1,2)}(u){-}C_{(k{+}1,l{-}2,2)}(u),\\
C_{(k{+}1,0,0)}(u)=u_1C_{(k,0,0)}(u){-}C_{(k{-}1,1,0)}(u){-}C_{(k{-}1,0,0)}(u),\\
C_{(0,l{+}1,0)}(u)=u_2C_{(0,l,0)}(u){-}C_{(2,l{-}1,0)}(u){-}C_{(0,l{-}1,0)}(u){-}C_{(1,l{-}1,2)}(u){-}C_{(1,l{-}2,2)}(u),\\
C_{(0,0,m{+}1)}(u)=u_3C_{(0,0,m)}(u){-}C_{(0,1,m{-}1)}(u){-}C_{(1,0,m{-}1)}(u){-}C_{(0,0,m{-}1)}(u),\\
C_{(k{+}1,0,m)}(u)=u_1C_{(k,0,m)}(u){-}C_{(k,1,m{-}2)}(u){-}C_{(k{-}1,1,m)}(u),\\
C_{(k,0,m{+}1)}(u)=u_3C_{(k,0,m)}(u){-}C_{(k{-}1,0,m{+}1)}(u){-}C_{(k{-}1,1,m{-}1)}(u){-}C_{(k,1,m{-}1)}(u)\\
\phantom{C_{(k,0,m{+}1)}(u)=}                    {-}C_{(k,0,m{-}1)}(u){-}C_{(k{+}1,0,m{-}1)}(u),\\
C_{(0,l{+}1,m)}(u)=u_2C_{(0,l,m)}(u){-}C_{(1,l{-}1,m{+}2)}(u){-}C_{(1,l{-}2,m{+}2)}(u){-}C_{(0,l{-}1,m)}(u)\\
\phantom{C_{(0,l{+}1,m)}(u)=}                    {-}C_{(2,l{-}1,m)}(u){-}C_{(1,l{+}1,m{-}2)}(u){-}C_{(1,l,m{-}2)}(u),\\
C_{(0,l,m{+}1)}(u)=u_3C_{(0,l,m)}(u){-}C_{(1,l{-}1,m{+}1)}(u){-}C_{(0,l{-}1,m{+}1)}(u){-}C_{(0,l{+}1,m{-}1)}(u)\\
\phantom{C_{(0,l,m{+}1)}(u)=}                    {-}C_{(0,l,m{-}1)}(u){-}C_{(1,l,m{-}1)}(u).\\
\end{gather*}

The lowest $C$-polynomials of $B_3$ were calculated explicitly and arranged in Table~\ref{c_poly_b3}.
\begin{table}[h]
\begin{center}
$\begin{array}{c}
\sharp 0
\\
\hline  \hline
\begin{array}{|l|r|r|r|r|r|r|r|r|r|r|r|r|r|r|}
   C_{(k, l,m)}(u)&1& u_1&u_2&u_1^2& u_3^2&u_1u_2 &u_1u_3^2&u_1^3&u_2^2  &u_2u_3^2&u_1^2u_2  \\ \hline
   C_{(0,0,0)}(u) &1&    &   &     &      &       &       &        &       &       &            \\ \hline
   C_{(1,0,0)}(u) &0&  1 &   &     &      &       &       &        &       &       &                \\ \hline
   C_{(0,1,0)} (u)&0&  0 & 1 &     &      &       &       &        &       &       &             \\ \hline
   C_{(2,0,0)} (u)&-6& 0 &-2 & 1   &      &       &       &        &       &       &        \\ \hline
   C_{(0,0,2)} (u)&-8& -4&-2 & 0   & 1    &       &       &        &       &       &       \\ \hline
   C_{(1,1,0)}(u) &24& 8 & 6 & 0   & -3   & 1     &       &        &       &       &       \\ \hline
   C_{(1,0,2)}(u) &0& -8 &-2 &-4   & 0    & -2    & 1     &        &       &       &           \\ \hline
   C_{(3,0,0)}(u)&-24&-15& -6& 0   & 3    &  -3   & 0     &    1   &       &       &       \\ \hline
   C_{(0,2,0)} (u)&12& 16& 8 & 4   & 0    & 4     &  -2   &    0   &   1   &       &          \\ \hline
   C_{(0,1,2)}(u)&-48&-20&-20& 0   & 6    &  -6   &0      &    0   &  -2   &   1   &          \\ \hline
   C_{(2,1,0)} (u)&0& 8  &-6 & 4   & 0    &  2    & -1    &    0   &  -2   &   0   &    1     \\ \hline
\end{array}
\end{array}$
\\[2ex]
$\begin{array}{c}
\sharp 1
\\
\hline  \hline
\begin{array}{|l|r|r|r|r|r|r|r|}
   C_{(k, l,m)}(u)&u_3& u_1u_3&u_2u_3 & u_3^3&u_1^2u_3 &u_1u_2u_3\\\hline
   C_{(0,0,1)}(u) & 1 &       &       &      &         &        \\\hline
   C_{(1,0,1)}(u) &-3 & 1     &       &      &         &            \\ \hline
   C_{(0,1,1)}(u) & 3 & -2    & 1     &      &         &         \\ \hline
   C_{(0,0,3)}(u) & -9& -3    &-3     & 1    &         &          \\ \hline
   C_{(2,0,1)}(u) & -3&-1     &-2     & 0    &  1      &        \\ \hline
   C_{(1,1,1)}(u)&3 0 & 12    & 8     &  -3  & -2      &    1    \\ \hline
\end{array}
\end{array}$
\end{center}
\caption{Lowest $C$-polynomials of $B_3$ split into two congruence classes $\#=0$ and $\#=1$.}\label{c_poly_b3}
\end{table}

As in the previous case, we can use the Weyl character formula
or generic recurrence relations for $S$-polynomials of $B_3$ valid for $k,l,m\geqslant 2$:
\begin{gather*}
\mathcal{S}_{(k{+}1,l,m)}(u)=u_1\mathcal{S}_{(k,l,m)}(u){-}\mathcal{S}_{(k,l{+}1,m{-}2)}(u){-}\mathcal{S}_{(k,l{-}1,m{+}2)}(u){-}\mathcal{S}_{(k{+}1,l{-}1,m)}(u)\\
\phantom{\mathcal{S}_{(k{+}1,l,m)}(u)=}        {-}\mathcal{S}_{(k{-}1,l{+}1,m)}(u){-}\mathcal{S}_{(k{-}1,l,m)}(u){-}\mathcal{S}_{(k{-}1,l{+}1,m)}(u){-}\mathcal{S}_{(k{-}1,l,m)}(u),\\
\mathcal{S}_{(k,l{+}1,m)}(u)=u_2\mathcal{S}_{(k,l,m)}(u){-}\mathcal{S}_{(k{+}1,l{-}1,m{+}2)}(u){-}\mathcal{S}_{(k{-}1,l{-}1,m{+}2)}(u){-}\mathcal{S}_{(k{+}1,l{-}2,m{+}2)}(u)\\
\phantom{\mathcal{S}_{(k,l{+}1,m)}(u)=}         {-}\mathcal{S}_{(k{+}1,l,m{-}2)}(u){-}\mathcal{S}_{(k{-}2,l{+}1,m)}(u){-}\mathcal{S}_{(k{-}1,l,m{+}2)}(u){-}\mathcal{S}_{(k{-}1,l{+}2,m{-}2)}(u)\\
\phantom{\mathcal{S}_{(k,l{+}1,m)}(u)=}         {-}\mathcal{S}_{(k{+}1,l{+}1,m{-}2)}(u){-}\mathcal{S}_{(k{-}1,l{+}1,m{-}2)}(u){-}\mathcal{S}_{(k{+}2,l{-}1,m)}(u),\\
\mathcal{S}_{(k,l,m{+}1)}(u)=u_3\mathcal{S}_{(k,l,m)}(u){-}\mathcal{S}_{(k{+}1,l{-}1,m{+}1)}(u){-}\mathcal{S}_{(k,l{-}1,m{+}1)}(u){-}\mathcal{S}_{(k{-}1,l,m{+}1)}(u)\\
\phantom{\mathcal{S}_{(k,l,m{+}1)}(u)=}         {-}\mathcal{S}_{(k{-}1,l{+}1,m{-}1)}(u){-}\mathcal{S}_{(k,l{+}1,m{-}1)}(u){-}\mathcal{S}_{(k{+}1,l,m{-}1)}(u){-}\mathcal{S}_{(k,l,m{-}1)}(u).
\end{gather*}

\subsection{$C$- and $S$-polynomials of $C_3$ in three real variables}
Low order $C$-polynomials of group $C_3$ are listed in Table~\ref{c_poly_c3}.
Higher-order polynomials can be obtained from the recurrence relations.
Generic recursions for $k,l,m\geqslant 2$ are:
\begin{gather*}
C_{(k{+}1,l,m)}(u)=u_1C_{(k,l,m)}(u){-}C_{(k,l{-}1,m{+}1)}(u){-}C_{(k{+}1,l{-}1,m)}(u){-}C_{(k{-}1,l{+}1,m)}(u)\\
\phantom{C_{(k{+}1,l,m)}(u)=}        {-}C_{(k{-}1,l,m)}(u){-}C_{(k,l{+}1,m{-}1)}(u),\\
C_{(k,l{+}1,m)}(u)=u_2C_{(k,l,m)}(u){-}C_{(k{+}1,l,m{-}1)}(u){-}C_{(k{-}1,l,m{+}1)}(u){-}C_{(k,l{-}1,m)}(u)\\
\phantom{C_{(k,l{+}1,m)}(u)=}        {-}C_{(k{+}1,l{+}1,m{-}1)}(u){-}C_{(k{-}2,l{+}1,m)}(u){-}C_{(k{-}1,l{-}1,m{+}1)}(u){-}C_{(k{-}1,l{+}1,m{-}1)}(u)\\
\phantom{C_{(k,l{+}1,m)}(u)=}        {-}C_{(k{+}2,l{-}1,m)}(u){-}C_{(k{+}1,l{-}1,m{+}1)}(u){-}C_{(k{+}1,l{-}2,m{+}1)}(u){-}C_{(k{-}1,l{+}2,m{-}1)}(u),\\
C_{(k,l,m{+}1)}(u)=u_3C_{(k,l,m)}(u){-}C_{(k,l{-}2,m{+}1)}(u){-}C_{(k{-}2,l,m{+}1)}(u){-}C_{(k,l{+}2,m{-}1)}(u)\\
\phantom{C_{(k,l,m{+}1)}(u)=}        {-}C_{(k,l,m{-}1)}(u){-}C_{(k{+}2,l{-}2,m{+}1)}(u){-}C_{(k{-}2,l{+}2,m{-}1)}(u){-}C_{(k{+}2,l,m{-}1)}(u).
\end{gather*}

Additional recursions:
\begin{gather*}
C_{(k{+}1,0,0)}(u)=u_1C_{(k,0,0)}(u){-}C_{(k{-}1,1,0)}(u){-}C_{(k{-}1,0,0)}(u),\quad k>1;\\
C_{(k{+}1,l,0)}(u)=u_1C_{(k,l,0)}(u){-}C_{(k,l{-}1,1)}(u){-}C_{(k{-}1,l,0)}(u)\\
\phantom{C_{(k{+}1,l,0)}(u)=}        {-}C_{(k{+}1,l{-}1,0)}(u){-}C_{(k{-}1,l{+}1,0)}(u),\quad k,l>1;\\
C_{(k,l{+}1,0)}(u)=u_2C_{(k,l,0)}(u){-}C_{(k{-}1,l,1)}(u){-}C_{(k,l{-}1,0)}(u){-}C_{(k{+}2,l{-}1,0)}(u){-}C_{(k{-}2,l{+}1,0)}(u)\\
\phantom{C_{(k,l{+}1,0)}(u)=}        {-}C_{(k{-}1,l{-}1,1)}(u){-}C_{(k{+}1,l{-}1,1)}(u){-}C_{(k{+}1,l{-}2,1)}(u),\quad k,l>2;\\
C_{(0,l{+}1,m)}(u)=u_2C_{(0,l,m)}(u){-}C_{(1,l,m{-}1)}(u){-}C_{(0,l{-}1,m)}(u){-}C_{(2,l{-}1,m)}(u){-}C_{(1,l{+}1,m{-}1)}(u)\\
\phantom{C_{(0,l{+}1,m)}(u)=}        {-}C_{(1,l{-}1,m{+}1)}(u){-}C_{(1,l{-}2,m{+}1)}(u),\quad l>2,\ m>1;\\
C_{(0,l,m{+}1)}(u)=u_3C_{(0,l,m)}(u){-}C_{(0,l{-}2,m{+}1)}(u){-}C_{(0,l,m{-}1)}(u){-}C_{(2,l{-}2,m{+}1)}(u)\\
\phantom{C_{(0,l,m{+}1)}(u)=}        {-}C_{(2,l,m{-}1)}(u){-}C_{(0,l{+}2,m{-}1)}(u),\quad l>2,\ m>1;\\
C_{(k{+}1,0,m)}(u)=u_1C_{(k,0,m)}(u){-}C_{(k{-}1,1,m)}(u){-}C_{(k,1,m{-}1)}(u){-}C_{(k{-}1,0,m)}(u),\quad k,m>1;\\
C_{(k,0,m{+}1)}(u)=u_3C_{(k,0,m)}(u){-}C_{(k{-}2,0,m{+}1)}(u){-}C_{(k,0,m{-}1)}(u){-}C_{(k{-}2,2,m{-}1)}(u)\\
\phantom{C_{(k,0,m{+}1)}(u)=}        {-}C_{(k{+}2,0,m{-}1)}(u){-}C_{(k,2,m{-}1)}(u),\quad k>2,\ m>1;\\
C_{(0,l{+}1,0)}(u)=u_2C_{(0,l,0)}(u){-}C_{(0,l{-}1,0)}(u){-}C_{(2,l{-}1,0)}(u){-}C_{(1,l{-}1,1)}(u){-}C_{(1,l{-}2,1)}(u),\;\; l>2;\\
C_{(0,0,m{+}1)}(u)=u_3C_{(0,0,m)}(u){-}C_{(0,0,m{-}1)}(u){-}C_{(2,0,m{-}1)}(u){-}C_{(0,2,m{-}1)}(u),\quad m>1.
\end{gather*}

Generic recursions for $S$-polynomials hold true when $k,l,m\geqslant 2$
\begin{gather*}
\mathcal{S}_{(k{+}1,l,m)}(u)=u_1\mathcal{S}_{(k,l,m)}(u){-}\mathcal{S}_{(k,l{-}1,m{+}1)}(u){-}\mathcal{S}_{(k{+}1,l{-}1,m)}(u){-}\mathcal{S}_{(k{-}1,l{+}1,m)}(u)\\
\phantom{\mathcal{S}_{(k{+}1,l,m)}(u)=}        {-}\mathcal{S}_{(k{-}1,l,m)}(u){-}\mathcal{S}_{(k,l{+}1,m{-}1)}(u),\\
\mathcal{S}_{(k,l{+}1,m)}(u)=u_2\mathcal{S}_{(k,l,m)}(u){-}\mathcal{S}_{(k{+}1,l,m{-}1)}(u){-}\mathcal{S}_{(k{-}1,l,m{+}1)}(u){-}\mathcal{S}_{(k{+}2,l{-}1,m)}(u)\\
\phantom{\mathcal{S}_{(k,l{+}1,m)}(u)=}        {-}\mathcal{S}_{(k{-}2,l{+}1,m)}(u){-}\mathcal{S}_{(k{-}1,l{-}1,m{+}1)}(u){-}\mathcal{S}_{(k{-}1,l{+}1,m{-}1)}(u){-}\mathcal{S}_{(k{+}1,l{+}1,m{-}1)}(u)\\
\phantom{\mathcal{S}_{(k,l{+}1,m)}(u)=}        {-}\mathcal{S}_{(k{+}1,l{-}1,m{+}1)}(u){-}\mathcal{S}_{(k{+}1,l{-}2,m{+}1)}(u){-}\mathcal{S}_{(k{-}1,l{+}2,m{-}1)}(u){-}\mathcal{S}_{(k,l{-}1,m)}(u),\\
\mathcal{S}_{(k,l,m{+}1)}(u)=u_3\mathcal{S}_{(k,l,m)}(u){-}\mathcal{S}_{(k,l{-}2,m{+}1)}(u){-}\mathcal{S}_{(k{-}2,l,m{+}1)}(u){-}\mathcal{S}_{(k,l{+}2,m{-}1)}(u)\\
\phantom{\mathcal{S}_{(k,l,m{+}1)}(u)=}        {-}\mathcal{S}_{(k{+}2,l{-}2,m{+}1)}(u){-}\mathcal{S}_{(k{-}2,l{+}2,m{-}1)}(u){-}\mathcal{S}_{(k{+}2,l,m{-}1)}(u){-}\mathcal{S}_{(k,l,m{-}1)}(u).
\end{gather*}

\begin{table}[h]
\begin{center}
$\begin{array}{c}
\sharp 0     \\
\hline  \hline
\begin{array}{|l|r|r|r|r|r|r|r|r|r|r|r|r|r|r|}
    C_{(k, l,m)}(u) &1 &u_2&u_1^2&u_1u_3& u_2^2&u_1^2u_2&u_3^2&u_1u_2u_3 &u_2^3  &u_2u_3^2    \\\hline
    C_{(0,0,0)} (u)   &1 &   &     &      &      &        &     &          &       &            \\\hline
    C_{(0,1,0)} (u)   &0 & 1 &     &      &      &        &     &          &       &          \\\hline
    C_{(2,0,0)} (u)   &-6&-2 &1    &      &      &        &     &          &       &         \\\hline
    C_{(1,0,1)} (u)   & 0&-2 &0    &  1   &      &        &     &          &       &            \\ \hline
    C_{(0,2,0)} (u)   &12& 8 & -4  &  -2  &   1  &        &     &          &       &            \\ \hline
    C_{(2,1,0)} (u)   & 0&-6 &  0  &   -1 &   -2 &    1   &     &          &       &            \\\hline
    C_{(0,0,2)} (u)   &-8& -8&  4  &   4  &   -2 &   0    & 1   &          &       &            \\\hline
    C_{(1,1,1)} (u)   & 0& 12& 0   &   -4 &   4  &   -2   &  -3 & 1        &       &       \\\hline
    C_{(0,3,0)} (u)   & 0& 9 & 0   &   3  &   6  &   -3   &  3  &-3        &1      &                     \\\hline
    C_{(0,1,2)} (u)   & 0&-18&  0  &  3   &   -12&    6   & 3   & 3        &-2     &1    \\\hline
\end{array}
\end{array}$
\\[3ex]
$\begin{array}{c}
\sharp 1    \\
\hline  \hline
\begin{array}{|l|r|r|r|r|r|r|r|r|r|r|r|r|r|rrrrrrrrrrrrrrrr}
       C_{(k, l,m)}(u) &u_1  &u_3& u_1u_2& u_1^3&u_2u_3&u_1^2u_3 &u_1u_2^2   &u_1u_3^2   &u_2^2u_3   &v_3^3\\\hline
       C_{(1,0,0)}   (u) & 1   &   &       &      &      &          &          &           &           &\\\hline
       C_{(0,0,1)}(u)    & 0   & 1 &       &      &      &          &          &           &           &      \\\hline
       C_{(1,1,0)}(u)    &-4   &-3 &   1   &      &      &          &          &           &           &\\\hline
       C_{(3,0,0)} (u)   &-3   &3  &   -3  &  1   &     &           &          &           &           &   \\ \hline
       C_{(0,1,1)} (u)   & 4   & 6 &   -2  &  0   &  1  &           &          &           &           &   \\ \hline
       C_{(2,0,1)}(u)    &  0  & -9&   0   &   0  &  -2 &  1        &          &           &           &   \\\hline
       C_{(1,2,0)}(u)    & 12  & -3&   9   &   -4 & -1  &   -2      & 1        &           &           &     \\\hline
       C_{(1,0,2)}(u)    &-12  & -6&   -6  &  4   &  -1 &  4        &-2        &1          &           &          \\\hline
       C_{(0,2,1)}(u)    & 0   & 27&    0  &  0   &  12 & -6        & 0        &-2         &1          &               \\\hline
       C_{(0,0,3)} (u)   & 0   &-27&    0  &  0   & -18 &  9        &0         & 6         &-3         &1    \\\hline
\end{array}
\end{array}$
\end{center}
\caption{Lowest $C$-polynomials of $C_3$ split into two congruence classes $\#=0$ and $\#=1$.}\label{c_poly_c3}
\end{table}

\section*{Acknowledgements}
Work supported in part by the Natural Sciences and Engineering Research Council of Canada, MITACS, and by the MIND Research Institute.
M.N. is grateful for the hospitality extended to her at the Centre de Recherches Math\'ematiques, Universit\'e de Montr\'eal, where a part of the work was carried out.
%%%%%%%%%%%%%%%%%%%%%%%%%%%%%%%%%%%%%%%%%%%%%%%%%%%%%%


\begin{thebibliography}{99}\small
\itemsep-0.5ex
%\medskip

\bibitem{Bourbaki}
Bourbaki~N.,
\emph{Lie groups and Lie algebras}, Chapters 1--3,
Springer-Verlag, Berlin~-- New York, 1989.

\bibitem{BMP}
Bremner~M.R., Moody~R.V., Patera~J.,
\emph{Tables of dominant weight multiplicities for representations of simple Lie algebras},
Marcel Dekker, New York, 1985. %, 340 pages.

%\bibitem{B}
%Bremner~M.R.,
%Fast computation of weight multiplicities,
%\textit{Journal of Symbolic Computation} \textbf{2} (1986), no. 4, 357--363.

\bibitem{Dunkl}
Dunkl~Ch., Xu~Yu.,
\emph{Orthogonal polynomials of several variables},
Cambridge University Press, New York, 2008.

%\bibitem{Dunkl2}
%Dunkl~Ch.,
%Some Orthogonal Polynomials in Four Variables,
%\textit{SIGMA} \textbf{4} (2008), 082, 9~pages, arXiv:0812.0063.

\bibitem{DunnLidl1980}
Dunn~K.B., Lidl~R.,
Multidimensional generalizations of the Chebyshev polynomials.~I,~II,
\textit{Proc. Japan Acad. Ser. A Math. Sci.} \textbf{56} (1980), no.~4, 154--159, 160--165.

\bibitem{dynkin}
Dynkin~E.B.,
Maximal subgroups of the classical groups,
\textit{Amer. Math. Soc. Transl. Series 2} {\bf 6} (1965), 245--378.

%\bibitem{FM}
%Filipov~S.N., Man'ko~V.I.,
%{Chebyshev polynomials and Fourier transform of $SU(2)$ irreducible representation character as spin-tomographic star-product kernel},
%(2009), arXiv:0905.4944v1.

\bibitem{GS}
Gaskell~R., Sharp~R.T.,
Generating functions for $G_2$ characters and subgroup branthing rules,
{\it J.~Math. Phys.} \textbf{22} (1981), 2736--2739.

\bibitem{HLP}
H\'akov\'a~L., Larouche~M., Patera~J.,
The rings of $n$-dimensional polytopes,
\textit{J.~Phys.~A: Math. Theor.} {\bf 41} (2008), 495202, 21 pages.

\bibitem{Heck}
Heckman G.J., Opdam E.M., Root systems and hypergeometric functions.~I,
{\it Compositio Math.} \textbf{64} (1987), 329--352.

\bibitem{HeckO}
Heckman G.J., Root systems and hypergeometric functions.~II,
{\it Compositio Math.} \textbf{64} (1987), 353--373.

\bibitem{HeckSchlich}
Heckman G., Schlichtkrull H.,
Harmonic analysis and special functions on symmetric spaces,
{\it Perspect.~Math.} \textbf{16}, Academic press, San Diego, 1994. %, 225 pages.



\bibitem{HrivnakPatera2009}%HP
Hrivn\'ak~J., Patera~J.,
On discretization of tori of compact simple Lie groups,
\textit{J.~Phys.~A: Math. Theor.} \textbf{42} (2009), 385208, 26~pages, arXiv:0905.2395.

\bibitem{HMP}
Hrivn\'ak J., Motlochov\'a L., Patera J.,
Two-dimensional symmetric and antisymmetric generalizations of sine functions,
\textit{J.~Math. Phys} {\bf 51} (2010), 073509, 13 pages, arXiv:0912.0241.

\bibitem{Humphreys1972}
Humphreys~J.E.,
\emph{Introduction to Lie algebras and representation theory},
Springer, New York, 1972.

\bibitem{KassMoodyPateraSlansky1990}
Kass~S., Moody~R.V., Patera~J., Slansky~R.,
\emph{Affine Lie algebras, weight multiplicities, and branching rules}, Vol.~1 and Vol.~2,
Los Alamos Series in Basic and Applied Sciences, University of California Press, Berkeley, 1990.

\bibitem{KlimykPatera2006}%KP1%KlimykPateraC
Klimyk~A., Patera~J.,
Orbit functions,
\textit{SIGMA} \textbf{2} (2006), 006, 60~pages, math-ph/0601037.

\bibitem{KlimykPatera2007-1}%KP2%KlimykPateraS
Klimyk~A., Patera~J.,
Antisymmetric orbit functions,
\textit{SIGMA} \textbf{3} (2007), 023, 83~pages, \mbox{math-ph/0702040}.

\bibitem{KlimykPatera2008}%KlimykPateraE
Klimyk~A., Patera~J.,
$E$-orbit functions,
\textit{SIGMA} \textbf{4} (2008), 002, 57~pages, arXiv:0801.0822.

\bibitem{KlimykPatera2007-3}%KP5%KlimykPateraSnExp
Klimyk~A., Patera~J.,
(Anti)symmetric multidimensional exponential functions and the corresponding Fourier transforms,
\textit{J.~Phys.~A: Math. Theor.} {\bf 40} (2007), 10473--10489, arXiv:0705.3572.

\bibitem{Koornwinder1-4}
Koornwinder~T.H.,
Orthogonal polynomials in two variables which are eigenfunctions of two algebraically independent partial differential operators. I--IV,
\textit{Indag. Math.} {\bf 36} (1974) 48--66, 357--381.

\bibitem{Koornwinder1993}
Koornwinder~T.H.,
Askey--Wilson polynomials for root systems of type $BC$,
in Hypergeometric functions on domains of positivity, Jack polynomials, and applications (Tampa, FL, 1991),
{\it Contemp. Math.}, {\bf 138}, Amer. Math. Soc., Providence, RI, 1992, 189--204.

\bibitem{Kowalski1982}
Kowalski~M.A.,
Orthogonality and recursion formulas for polynomials in $n$ variables,
\textit{Siam J. Math. Anal.} {\bf 13}, (1982) no. 2, 316--323.

\bibitem{Lapointe1}
Lapointe~L., Lascoux A., Morse~J.,
Determinantal expressions for Macdonald polynomials,
\textit{Internat. Math. Res. Notices} {\bf 1998} (1998), no.~18, 957--978.

\bibitem{Lapointe2}
Lapointe~L., Lascoux A., Morse~J.,
Determinantal expression and recursion for Jack polynomials,
\textit{Electron J. Combin.} \textbf{7} (2000), Note~1, 7 pages.

\bibitem{Lassalle}
Lassalle~M.,
 A short proof of generalized Jacobi--Trudi expansions for Macdonald polynomials,
 in Jack, Hall--Littlewood and Macdonald polynomials,
{\it Contemp. Math.}, {\bf 417}, Amer. Math. Soc., Providence, RI, 2006, 271--280, math/0401032.

\bibitem{Lassalle2006}
Lassalle~M., Schlosser~M.J.,
Inversion of the Pieri formula for Macdonald polynomials,
\textit{Adv. Math.} \textbf{202} (2006), 289--325, math/0402127.

\bibitem{LassalleSchlosser2010}
Lassalle~M., Schlosser~M.J.,
Recurrence formulas for Macdonald polynomials of type~$A$,
\textit{J.~Algebraic Combin.} \textbf{32} (2010), 113--131, arXiv:0902.2099.

\bibitem{LY}
Li~H.,  Xu~Y.,
Discrete Fourier analysis on fundamental domain of $A_d$ lattice and on simplex in $d$-variables,
\textit{J.~Fourier Anal. Appl.} {\bf 16} (2010),  383--433, arXiv:0809.1079.

\bibitem{Lidl}
Lidl~R., Wells~Ch.,
Chebyshev polynomials in several variables,
\textit{J. Reine Angew. Math.} \textbf{255} (1972), 104--111.

\bibitem{LP}
Lemire~F.W., Patera~J.,
Congruence number, a generalisation of $SU(3)$ triality,
\textit{J.~Math.~Phys.} {\bf 21}  (1980), 2026--2027.

\bibitem{Mac1}
Macdonald I.G.,
\textit{Symmetric functions and Hall polynomials}, 2nd ed.,
Oxford, Oxford Univ. Press, 1995.

\bibitem{Mac2}
Macdonald I.G.,
A new class of symmetric functions,
{\it Publ. I.R.M.A. Strasbourg}, 372/S-20, Actes 20 (1988),  S\'eminaire Lotharingien, 131--171.

\bibitem{Mac3}
Macdonald I.G.,
Orthogonal polynomials associated with root systems,
{\it S\'em. Lothar. Combin.} \textbf{45} (2000), Art.~B45a,  40~pages.

\bibitem{McP}
McKay~W.G., Patera~J.,
\emph{Tables of dimensions, indices, and branching rules for representations of simple Lie algebras},
Marcel Dekker, New York, 1981. %), 317 pages.

\bibitem{MP82}
Moody~R.V., Patera~J.,
Fast recursion formula for weight multiplicities,
\textit{Bull. Amer. Math. Soc.} {\bf 7}  (1982), 237--242.

\bibitem{MoodyPatera2010}
Moody~R.V., Patera~J.,
Cubature formulae for orthogonal polynomials in terms of elements of finite order of compact simple Lie groups,
\emph{Advances in Applied Mathematics} \textbf{47} (2011), 509--535, arXiv:1005.2773.

\bibitem{MP87}
Moody~R.V., Patera~J.,
Computation of character decompositions of class functions on compact semisimple Lie groups,
\textit{Math. Comp.} {\bf 48} (1987), 799--827.

\bibitem{MP06}
Moody~R.V., Patera~J.,
Orthogonality within the families of $C$-, $S$-, and $E$-functions of any compact semisimple Lie group,
\textit{SIGMA} {\bf 2} (2006), 076, 14~pages, math-ph/0611020.

\bibitem{MMP}
Moody~R.V., Motlochova~L., Patera~J.,
New families of Weyl group orbit functions,
in preparation.

\bibitem{NPT}
Nesterenko~M., Patera~J., Tereszkiewicz A.,
Orbit functions of $SU(n)$ and Chebyshev polynomials,
\textit{Proceedings of the 5th Workshop "Group Analysis of Differential Equations \& Integrable Systems''} (2010), 133--151, arXiv:0905.2925.

\bibitem{NPST}
Nesterenko~M., Patera~J., Szajewska~M., Tereszkiewicz A.,
Orthogonal polynomials of compact simple Lie groups: branching rules for polynomials,
{\it J.~Phys.~A: Math. Theor.} (2010), V. 43, 495207, 27 pages, arXiv:1007.4431.

\bibitem{Opdam}
Opdam~E.M.,
Harmonic analysis for certain representations of graded Hecke algebras,
\textit{Acta Math.} {\bf 175} (1995), 75--121.

\bibitem{Patera}%patera2005
Patera~J.,
Compact simple Lie groups and their $C$-, $S$-, and $E$-transforms,
\textit{SIGMA} {\bf 1} (2005), 025, 6~pages, math-ph/0512029.

\bibitem{PS}
Patera J., Sharp R.T.,
Generating functions for characters of group representations and their applications,
in Group Theoretial Methods in Physics, \textit{Springer Lecture Notes in Physics}, {\bf 79}, Springer, 1977, 175--183.

\bibitem{Rivlin1974}
Rivlin~T.J.,
\emph{The Chebyshev polynomials}, Wiley, New York, 1974.

\bibitem{suetin}
Suetin~P.K.,
\textit{Orthogonal polynomials in two variables},
Amsterdam, Gordon and Breach, 1999.

\bibitem{vanDiejen1}
van Diejen~J.F., Lapointe~L., Morse~J.,
Determinantal construction of orthogonal polynomials associated with root systems,
\textit{Compositio Math.} \textbf{140}, (2004), 255--273.

\bibitem{vanDiejen2}
van Diejen~J.F., Lapointe~L., Morse~J.,
Bernstein--Szego polynomials associated with root systems,
\textit{Bull. London Math. Soc.} \textbf{39}, (2007), 837--847.

\bibitem{vanDiejen3}
van Diejen~J.F., Emsiz~E.,
Pieri formulas for Macdonald’s spherical functions and polynomials,
to appear in \textit{Math. Z.} (2010), 11 pages, arXiv:1009.4482.

\bibitem{VilenkinKlimyk1995}
Vilenkin N.Ja., Klimyk A.U.,
\textit{Representations of Lie groups and special functions: recent advances},
Dordrecht, Kluwer, 1995.


\end{thebibliography}
\end{document}